%% file: Erratum.tex
\documentclass[twocolumn,fleqn]{openjournal}
\usepackage{times}
\usepackage{listings}
\usepackage{subfigure}
\usepackage{graphicx}
\usepackage{xspace}
\usepackage{amssymb}
\usepackage{amsmath}
\usepackage{hyperref}
\usepackage{rotating}
\usepackage{placeins}
\usepackage{natbib}
\usepackage{multirow}
\usepackage{float}
\usepackage{orcidlink}

\usepackage{xcolor}
\usepackage{textgreek}
\usepackage[utf8]{inputenc}
\usepackage[english]{babel}
\makeatletter
\providecommand{\sf@counterlist}{}
\makeatother
\usepackage[caption=false,subrefformat=parens]{subfig}
\hypersetup{
    unicode, 
    colorlinks=true,
    linkcolor=linkcolor,
    citecolor=linkcolor,
    filecolor=linkcolor,
    urlcolor=linkcolor,
}
\definecolor{linkcolor}{rgb}{0.0,0.3,0.5}

\renewcommand{\vec}[1] {\boldsymbol{#1}}

\shorttitle{Intrinsic alignment of disks and ellipticals}
\shortauthors{van Heukelum \& Chisari}

\begin{document}

\journalinfo{The Open Journal of Astrophysics}

\title{Intrinsic alignment of disks and ellipticals across hydrodynamical simulations}

\author{\vspace{-1.4cm}M. L. van Heukelum\,\orcidlink{0009-0008-3780-1617}$^{1,\ast}$}
\email{$^\ast$m.l.vanheukelum@uu.nl}
\author{N. E. Chisari\orcidlink{0000-0003-4221-6718}$^{1,2}$}

\affiliation{$^1$Institute for Theoretical Physics, Utrecht University, Princetonplein 5, 3584 CC, Utrecht, the Netherlands}
\affiliation{$^2$Leiden Observatory, Leiden University, P.O. Box 9513, 2300 RA Leiden, the Netherlands}

\begin{abstract}
The correlations between the positions and shapes of galaxies, i.e. intrinsic alignments, have been measured in many observational studies and hydrodynamical simulations.
The position-shape correlation measurements of disk galaxies with varying methodologies, samples and hydrodynamical simulations are inconsistent in amplitude and sign.
This work compares the correlations of disk and elliptical shapes around all galaxy positions and disk shapes around the positions of ellipticals at $z=0$ and $z=1$ for two different shape definitions in TNG300, Horizon-AGN and EAGLE for multiple morphological definitions in a consistent way.
All types of signals are positive and robust in TNG300 and EAGLE and positive or null in Horizon-AGN.
A re-weighting of the ellipticals around all galaxies correlations in TNG300, according to the underlying stellar mass distributions of the samples, suggests that stellar mass is the driving factor determining the amplitude of the correlations.
The exception to this is the negative correlation of disks around ellipticals in Horizon-AGN. This arises for reduced shapes, which down-weight the outskirts of galaxies, at $z=1$, when disks are identified via a threshold in $|v/\sigma|$, the rotational velocity over the velocity dispersion.
\end{abstract}

\maketitle

\section{Introduction}
\label{sec:intro}

Cosmology concerns itself with measuring, modelling and understanding the largest scales in the Universe.
The current best-describing cosmological model is the $\Lambda$ cold dark matter ($\Lambda$CDM) model, which assumes dark energy, cold dark matter and baryonic matter as the main components of our universe.
The energy densities of these components can be measured in various ways, of which weak gravitational lensing, the statistical correlation of the deformation of galaxy shapes due to the bending of light by matter, is a promising probe of $\Omega_m$, the matter density, and $\sigma_8$, which quantifies the amplitude of the density fluctuations in the universe in sphere of radius $8\,h^{-1}{\rm Mpc}$.
As Stage III surveys, Kilo Degree Survey (KiDS) \citep{KiDS}, Dark Energy Survy (DES) \citep{DES} and Hyper Suprime-Cam Subaru Strategic Program (HSC) \citep{HSC}, come to a close and the era of precision cosmology from Stage IV surveys such as Euclid \citep{Laureijs11} and the Legacy Survey of Space and Time at the NSF-DOE Vera C. Rubin Observatory (LSST) \citep{Ivezic19} begins, understanding the systematic uncertainties in weak lensing becomes more important.

One such contaminant in weak lensing is the intrinsic alignment (IA) of galaxies, which is a spatial correlation between galaxy positions and shapes that cannot be attributed to gravitational shear, but is instead caused by the (gravitational) interactions of galaxies with the surrounding large-scale structure. 
Apart from being able to model and mitigate the effects of IA on the cosmic shear, the alignment signal itself also carries cosmological information \citep[e.g.][]{Chisari_2013,Vlah_2021,Kogai_2021,Saga_2024} and the study of IA can lead to better understanding of galaxy evolution \citep{Kiessling_2015}.

One way to gain understanding of galaxy evolution is through the measurement of IA in hydrodynamical simulations, which allow us to study galaxies in ways that observations do not (yet) allow. 
For a decade, IA has been studied in hydrodynamical simulations to gain insight on the correlation between stellar mass, morphology, luminosity, AGN feedback, among other factors, and the amplitude of IA \citep[e.g][]{Tenneti_2014,Chisari_2015,Shao_2016,Tenneti_2017,Chisari_2018,Shi_2021}.
In general, the IA amplitude is always highest for elliptical, high mass and luminous galaxies \citep[e.g.][]{Chisari_2016,Tenneti_2016,Samuroff_2021}. 
Furthermore, the redshift evolution of the IA correlation functions has also been studied, which is crucial for its mitigation in weak lensing studies \citep[e.g.][]{welker2017caughtrhythmiicompetitive,Bhowmick_2019,Bate_2019}.

One of the open problems in this field is the alignment of disks, which might be the majority of weak lensing sources for Stage IV surveys.
Disk alignments have been measured by various studies in differing simulations and for a variety of morphological definitions and methodologies, with conflicting positive (radial), negative (tangential) and null detections.
\cite{Tenneti_2016} found positive disk around elliptical ellipticity-direction alignments in MassiveBlack-II \citep{Khandai_2015} and Illustris \citep{Genel2014,Vogelsberger2014}, when splitting galaxies via their bulge-to-total ratio.
\cite{Hilbert_2017} also found a positive disk alignment in Illustris, mimicking observations and selecting disks by their Bayesian Photometric redshift (BPZ) spectral type.
The alignment amplitude obtained from modelling the correlation functions with the nonlinear alignment model (NLA) at $z>0.5$ yield positive alignments in Illustris, MassiveBlack-II and TNG300 \citep{nelson2021illustristngsimulationspublicdata} for blue (disk) galaxies in the work of \cite{Samuroff_2021}.
When measuring the 3D power spectrum for galaxy alignments, \cite{Shi_2021} find a disk alignment consistent with zero in TNG300, when defining disks using $\kappa_{\mathrm{rot}}$ (see Eq. \ref{eq: kappa_rot}).
They do find a 'spin flip' signal, where the spin, i.e. the direction of the angular momentum, of low-mass disks is parallel to the filament and high mass disks are aligned at $z=0.3,0.5$.
In the newest TNG simulation, MTNG740, \cite{Delgado_2023} measure a positive disk IA signal for disks defined by $\kappa_{\mathrm{rot}}$ and sSFR.
\cite{Chisari_2015} found a negative disk alignment when measuring the correlations of disks around ellipticals, defined by $v/\sigma$ (see Sec. \ref{sec:md selections} for measurement method), in Horizon-AGN \citep{Dubois_2014} for reduced shapes, which are also found at higher redshift in \cite{Chisari_2016}.
\cite{Dubois_2014} measures the alignment of galaxies in Horizon-AGN and its dependence on various properties, e.g. $v/\sigma$, sSFR and colour, finding a positive spin-filament alignment for disks.
\cite{Kraljic_2020} measures the alignment between galaxy spin and the filaments of the matter distribution and finds that in SIMBA \citep{Dav__2019}, the alignment signal flips between high and low stellar mass samples, for both high and low $v/\sigma$ populations of galaxies.
A positive alignment between the spin of blue galaxies and the filament is also found by \cite{welker2017caughtrhythmiicompetitive} in Horizon-AGN, who also looked extensively into satellite alignments.
In EAGLE \citep{Schaye2015EAGLE,Crain2015EAGLE}, \cite{Velliscig_2015} measured the 3D alignment of galaxies based on their axis ratios and found a positive disk signal.

In observations, most disk alignments are found to be consistent with zero.
Using the SDSS Main sample, \citet{Mandelbaum_2006} measured no significant blue alignment for blue galaxies around the full sample, but did find the red and blue alignments consistent in the highest luminosity bin, which is attributed to the fact that these galaxies are almost red.
Later, \citet{Mandelbaum_2011} also found no alignment signal for selection of WiggleZ Dark Energy Survey galaxies that overlaps with SDSS, constraining the blue alignment in a model-dependent way.
\citet{Johnston2019} performed the measurement of intrinsic alignments of blue galaxies in KiDS+GAMA, leading to tighter constraints of the lack of alignment of blue galaxies.
Recently, \cite{georgiou2025}, who measured the disk signal in the KiDS-1000 bright sample for both blue and $n_s<2.5$ galaxies, also found no significant alignment for this population of galaxies.

Comparison between these studies is challenging due to varying samples, simulations and methodology for galaxy shapes and galaxy alignment measurements.
Therefore, this work aims to conduct a consistent comparison between three widely used hydrodynamical simulations, TNG300, EAGLE and Horizon-AGN, for multiple morphological definitions, to shed some light on the robustness of the disk alignment signal across morphological definitions and simulations.
As the scales studied are very non-linear and therefore difficult to model, this work places the focus on obtaining qualitative conclusions.

This work is structured as follows. 
Section \ref{sec:measurements and data} focusses on the simulation data (Sec. \ref{sec:md simulations}); measurement methodologies of galaxy shapes (Sec. \ref{sec:md shapes}); projected correlation functions (Sec. \ref{sec:md correlation functions}) and the variables used in the morphological definitions, which are used to select the samples (Sec. \ref{sec:md selections}).
Section \ref{sec:results} presents the results of the distributions of the morphological variables (Sec. \ref{sec:r distributions}); measurement of IA correlations of disks and ellipticals around all galaxies (Sec. \ref{sec:r AB}); disks around ellipticals (Sec. \ref{sec:r DxE}); the disk signal in Horizon-AGN (Sec. \ref{sec:r Disks}) and the influence of the stellar mass distributions on the ellipticals around all galaxies measurements (Sec. \ref{sec:r Mass}).
Section \ref{sec:discussion and outlook} provides discussion and an outlook for future research and the conclusions are summarised in Section \ref{sec:conclusions}.

\section{Measurements and Data}
\label{sec:measurements and data}

\subsection{Simulations}
\label{sec:md simulations}
Throughout this work, we use the data from three hydrodynamical simulation suites: IllustrisTNG \citep{nelson2021illustristngsimulationspublicdata} and EAGLE \citep{Schaye2015EAGLE,Crain2015EAGLE}, which are publicly available, and Horizon-AGN \citep{Dubois_2014}.
We chose these simulations because they are most widely used when studying intrinsic alignments, allowing for comparison to previous works.
Their resolutions and boxsizes allow for large enough samples of a representative group of galaxies, i.e. not just massive ones, but milky-way-like as well, with a enough signal-to-noise for a robust detection of intrinsic alignment correlation functions at non-linear scales.
These simulations are run using a variety of parameters, codes and models, which are described briefly below and summarised in Table \ref{table:sim stats}.
For more details, see the cited papers.

\begin{table*}
\caption{Parameter information of each simulation used in this work, including cosmology, boxsize, number of dark matter (DM) particles, DM particle mass, numerical scheme, AGN feedback model modes and level of calibration.}
    \centering
    \begin{tabular}{cccccccc}
        Simulation &Cosmology& Boxsize & Number of & DM particle & Numerical scheme  & AGN feedback & Calibration  \\
         & & &  DM particles &  mass &   &  &  level \\
        \hline
         TNG300 & Planck 2016 & 205 $\mathrm{cMpc}/h$ & $2500^3$ & $5.9\times10^7\mathrm{M_\odot}$ & moving mesh & thermal and kinetic & high \\\hline
         TNG100-1& Planck 2016 & 75 $\mathrm{cMpc}/h$ & $1820^3$ & $7.5\times10^6\mathrm{M_\odot}$ & moving mesh & thermal and kinetic & high \\\hline
         TNG100-2& Planck 2016 & 75 $\mathrm{cMpc}/h$ & $910^3$ & $6.0\times10^7\mathrm{M_\odot}$ & moving mesh & thermal and kinetic & high \\\hline
         EAGLE & Planck 2014& 67.77 $\mathrm{cMpc}/h$ & $1504^3$&$6.6\times10^6\mathrm{M_\odot}$ & smoothed particle hydrodynamics& thermal & low \\ \hline
         Horizon-AGN & WMAP7& 100 $\mathrm{cMpc}/h$ & $1024^3$ & $8\times10^7\mathrm{M_\odot}$ &adaptive mesh refinement & thermal and kinetic & low \\
    \end{tabular}
    
    \label{table:sim stats}
\end{table*}

\subsubsection{IllustrisTNG}
The IllustrisTNG project contains simulations with multiple boxsizes and resolutions, of which TNG300-1 is used throughout this work, and TNG100-1 and TNG100-2 are used in Appendix \ref{app:resolution} to gauge the influence of boxsize and resolution effects on the results.
The volume of TNG300 is a box with length $205\mathrm{cMpc}/h$ on each side, containing $2500^3$ dark matter particles of mass $m_{\mathrm{DM}}=5.9\times 10^7 \mathrm{M_\odot}$, which are evolved using the AREPO code \citep{Springel_2010} from $z=127$ to $z=0$.
The cosmological parameters are assumed to be: $\Omega_{\mathrm{m},0} = 0.3089$, 
$\Omega_{\mathrm{b},0} = 0.0486$, 
$\sigma_8 = 0.8159$, 
$n_\mathrm{s} = 0.9667$ and $h=0.6774$ \citep{Planck2016}.
Physical processes that are not resolved are governed by sub-grid models, which are described in \citet{Pillepich_2017_methods} and \citet{Weinberger_2016} and include procedures for radiative cooling and heating; star formation; stellar evolution and feedback; chemical enrichment and black hole accretion, mergers and AGN feedback with thermal (quasar) and kinetic (wind) modes.
Haloes, sub-haloes and galaxies are found and defined using the Friends-of-Friends (FoF) and SUBFIND algorithms \citep{SUBFIND_Springel_2001}, which group dark matter particles and then determine the gravitationally bound (sub-)structures.
The IllustrisTNG simulations have been calibrated to the $z=0$ galaxy stellar mass function; stellar mass - halo mass relation; stellar mass - stellar size relation; stellar mass - BH mass relation and total gas mass content (in $r_{500}$) in massive galaxy groups and the shape of the cosmic star formation rate density at $z\leq10$.

\subsubsection{EAGLE}
The EAGLE simulation suite also contains multiple boxsizes and resolutions, of which the largest volume, with box length $100\mathrm{cMpc}=67.77\mathrm{cMpc}/h$, is used in this work.
It contains $1504^3$ dark matter particles with mass $m_{\mathrm{DM}}=6.6\times 10^6 \mathrm{M_\odot}$ and uses the Planck cosmological parameters: $\Omega_{\mathrm{m}} = 0.307$,
$\Omega_{\mathrm{b}} = 0.04825$, 
$\sigma_8 = 0.8288$, $n_\mathrm{s} = 0.9611$ and $h=0.6777$ \citep{Planck2014}.
The simulations are run using a modified version of the smoothed particle hydrodynamics code GADGET 3 \cite{springel2005cosmological}.
As for TNG, sub-grid physics models are added to describe unresolved physics: element-by-element radiative cooling and heating; star formation and feedback; stellar mass loss and the accretion and mergers of black holes, along with thermal AGN feedback.
In EAGLE, the sub-grid model parameters are calibrated to reproduce the galaxy stellar mass function and the galaxy sizes at $z=0$.
Haloes, sub-haloes and galaxies are found using the FoF and SUBFIND algorithms as in TNG.

\subsubsection{Horizon-AGN}

The Horizon-AGN simulation \citep{Dubois_2014} has a volume of $100^3[\mathrm{cMpc}/h]^3$ and $1024^3$ dark matter particles of mass $m_{\mathrm{DM}}=8\times 10^7 \mathrm{M_\odot}$.
Horizon-AGN uses the Wilkinson Microwave Anisotropy Probe year 7 \citep[WMAP7]{Komatsu_2011} cosmological parameters: $\Omega_{\mathrm{m}} = 0.272$,
$\Omega_{\mathrm{b}} = 0.045$, 
$\sigma_8 = 0.81$, $n_\mathrm{s} = 0.967$ and $h=0.704$.
This simulation is run using the adaptive mesh refinement code RAMSES \citep{teyssier2002cosmological}, which evolves gas cells that are resolved further depending on the number of dark matter particles.
Similar to the other two simulations, sub-grid models govern unresolved physics including gas cooling and heating; stellar winds; star formation; supernova feedback; black hole accretion and AGN feedback in kinetic (radio) and thermal (quasar) modes.
Galaxies are found using the AdaptaHOP algorithm \citep{Aubert2004}, which uses the distribution of stellar particles to identify galaxies with a minimum of 50 stellar particles.
Horizon-AGN has been calibrated to reproduce the $M-\sigma$ relation at $z=0$.

\subsection{Shapes of galaxies}
\label{sec:md shapes}
For consistency, the shapes used in this work have all been calculated using the same method for each simulation.
In this entire work, the projected shapes are calculated using the simple, Eq. \ref{eq: inertia tensor}, or reduced inertia tensor, Eq. \ref{eq: reduced inertia tensor}, defined by:
\begin{align}
    I^{k,l}_s &= \frac{1}{M_s} \sum^{N_s}_{n=1} m_n r^k_n r^l_n \label{eq: inertia tensor}\\
    \tilde{I}^{k,l}_r &= \frac{1}{M_s} \sum^{N_s}_{n=1} m_n \frac{r^k_n r^l_n}{r^2} \label{eq: reduced inertia tensor}
\end{align}
where $k,l$ run from $0$ to $1$ and denote the $x,y$ (or $z$) components of vector $\vec{r_n}$, depending on the projection axis (line-of-sight), which is chosen to be $z$ in this work.
$\vec{r_n}$ is the position of each particle $n$ in galaxy $s$ relative to the centre of mass of the galaxy ($\vec{r_n} = \vec{x_n}-\vec{x_s}$), with $n$ running from $1$ to $N_s$.
The mass-weighted centre of mass of the stellar particles for each galaxy ($s$) is calculated as follows:
\begin{equation}
    \vec{x}_{s} = \frac{1}{M_s} \sum^{N_s}_{n=1} m_n \vec{x}_n
    \label{eq: COM}
\end{equation}
where $M_s$ is the galaxy stellar mass ($\sum_n m_n$) and $\vec{x_n}$ and $m_n$ are the stellar particle positions and masses.
In TNG300, the `wind-particles', which are actually gas particles, have been omitted from the stellar mass.

The eigenvectors of the simple (reduced) inertia tensor, Equation \ref{eq: inertia tensor} (\ref{eq: reduced inertia tensor}), are the unit vectors that denote the directions of the semi-major and semi-minor axes in the galaxy shape.
For the simple inertia tensor, the axis lengths are given by the square root of the eigenvalues and named, $a$ (semi-major axis length) and $b$ (semi-minor).

\subsection{Projected correlation functions}
\label{sec:md correlation functions}

To match what is measured in observations, we use projected statistics of galaxy alignments.
In this work, we use two statistics to quantify the alignments of galaxy shapes around galaxy positions: $\tilde{\xi}_{g+,2}$ (the quadrupole) and $w_{g+}$ (only in Appendix \ref{app:wg+}).
The correlation functions are estimated using the Landy-Szalay estimators \citep{LandySzalay}, which use a shape sample, $S$, a position sample, $D$ and a random sample, $R$, to estimate the correlation of galaxy shapes and positions, separated by $\vec{r}$.
Both statistics use the same estimator, but are binned in different ways: either in $(r_p,\Pi)$, where $r_p$ is the length of $\vec{r}$, projected along the line-of-sight, $\Pi$, or in $(r,\mu_r)$, where $r$ is the length of $\vec{r}$ and $\mu_r=\Pi/r$, the cosine of the angle between $\vec{r}$ and $\Pi$.
Regardless of the binning, the modified Landy-Szalay estimator used in this work \citep{Mandelbaum_2011} is defined as:
\begin{align}
    \xi_{g,+}(r_p,\Pi) &= \frac{S_+D - S_+R}{RR}\label{eq:xi_g+}\\
    S_+D &= \sum_{\{i,j\}}^{(r_p,\Pi)} w_i w_j \frac{e_{+}(j|i)}{2\mathcal{R}}\label{eq:S+D}\\
    (e_{+},e_{\times}) &= \frac{1-q^2}{1+q^2}[\cos(2\phi)\sin(2\phi)]
    \label{eq:e+}
\end{align}
Here, $(r,\mu_r)$ can be substituted for $(r_p,\Pi)$.
As shown in Equations \ref{eq:xi_g+} - \ref{eq:e+}, the correlation functions can be estimated by first finding all galaxy pairs $\{i,j\}$ between the position and shape sample in an $(r_p,\Pi)$ bin.
For each pair, the tangential/radial component of the ellipticity of the shape galaxy $j$ (given position $i$) can be obtained using Equation \ref{eq:e+}, where $q$ is the axis ratio of the projected shape, $q=b/a$, as defined in Section \ref{sec:md shapes} and $\phi$ is the angle between the separation vector $\vec{r}$ and the semi-major axis direction (also defined in Section \ref{sec:md shapes}).
A summation of $e_+$ over all galaxy pairs within an $(r_p,\Pi)$ or $(r,\mu_r)$ bin, is then divided by the responsivity factor $\mathcal{R}$, defined as $\mathcal{R}=1-\langle e^2 \rangle$, when unweighted, with $e=\frac{1-q^2}{1+q^2}$, to obtain $S_+D$ (Eq. \ref{eq:S+D}).
Each galaxy in the shape and position sample has weight $w_j,w_i$, respectively, which are set to $1$ in all sections, except in Section \ref{sec:r Mass}.
If the weights are not unity, $\mathcal{R}$ is defined as $\mathcal{R}=\frac{\sum_j{w_j(1-e^2/2)}}{\sum_j{w_j}}$, in line with \citet{Jarvis_2003}.
Equivalently to $S_+D$, $S_+R$ is the summation over galaxy ellipticities around random points. This term is be assumed to be zero in this work, following \citet{Chisari_2015}.

The $RR$ sample counts the expected number of galaxy pairs in each bin for a random sample, which is equivalent to the number of galaxies in the shape sample multiplied by the number in the position sample, multiplied by the volume of the bin.
This can be calculated analytically for $(r_p,\Pi)$ and $(r,\mu_r)$ bins.
The $(r,\mu_r)$ bins are in the form of a spherical cap shell slice, which is defined by the difference between the volume of two spherical cap slices with differing heights defined by the boundaries of $\mu_r$.

\subsubsection{Projected correlation function $w_{g+}$}

The most widely used statistic to measure the projected shape-position correlation functions of IA is $w_{g+}$.
We can obtain $w_{g+}$ by integrating $\xi_{g+}(r_p,\Pi)$ (Eq. \ref{eq:xi_g+}) over the line of sight $\Pi$:
\begin{equation}
    w_{g+}(r_p) = \int_{-\Pi_{\text{max}}}^{\Pi_{\text{max}}} \xi_{g+}(r_p, \Pi) \text{d}\Pi
    \label{eq:w_g+}
\end{equation}
where $\Pi_{max}$ is the chosen integration limit, set to half the boxsize in this work, equating to $102.5\mathrm{Mpc}/h$ for TNG300.
As the quadrupole (see below) has a higher signal-to-noise than $w_{g+}$ \citep{singh2024increasingpowerweaklensing}, we have chosen to show only the quadrupole in Section \ref{sec:results}, but some of the $w_{g+}$ measurements are added to Appendix \ref{app:wg+} for comparison to other works.
Any quadrupole measurements in this work are also available as $w_{g+}$ measurements upon request.

\subsubsection{Multipoles}
\citet{singh2024increasingpowerweaklensing} introduced a new estimator for the measurement of projected alignment statistics, which has increased signal-to-noise compared to $w_{g+}$ and expresses the correlations in terms of their multipole moments:
\begin{align}
    \tilde{\xi}_{kl}^{\ell,s_{ab}} (r) &= \frac{2\ell + 1}{2}\frac{(\ell - s_{ab})!}{(\ell + s_{ab})!}\int \text{d} \mu_{r}L^{\ell,s_{ab}}(\mu_r)\xi_{jk}(r, \mu_{r})
\end{align}
where $k$ and $l$ can be $g$ or $+$; our focus in this work lies with $k=g$ and $l=+$.
The previously defined $\xi_{g+}(r,\mu_r)$ Eq. \ref{eq:xi_g+} is inserted into the integrand, along with the appropriate associated Legendre polynomials, where $s_{ab}=\ell=2$ are needed for $\xi_{g+}$.

Filling in the terms, we get:
\begin{equation}
 \tilde{\xi}_{g+}^{2,2} (r) = \frac{5}{48}\int \text{d} \mu_{r}L^{2,2}(\mu_r)\xi_{g+}(r, \mu_{r}) .
\end{equation}
where $\tilde{\xi}_{g+}^{2,2} (r)$, will be abbreviated to $\tilde{\xi}_{g+,2} (r)$ and called 'quadrupole' throughout the rest of this work.

For the measurements of both $w_{g+}$ and $\tilde{\xi}_{g+,2}$ the code {\sc{MeasureIA}}\xspace\footnote{https://github.com/MarloesvL/measure\_IA}\footnote{doi:10.5281/zenodo.17252215} was developed and used.
{\sc{MeasureIA}}\xspace is publicly available and has been validated against {\sc{halotools}}\xspace\footnote{ https://github.com/astropy/halotools} \citep{halotools} and {\sc{TreeCorr}}\xspace\footnote{https://github.com/rmjarvis/TreeCorr} \citep{treecorr} for $w_{g+}$ and $w_{gg}$ measurements, including covariance.

\subsubsection{Covariance}

The covariance is estimated using the jackknife method.
In this method, the simulation volume is divided into sub-volumes of equal size, and the correlation functions are measured for the entire volume, omitting one sub-volume at a time.
This process is sensitive to the number of sub-volumes, or jackknife regions, $N_{\mathrm{sub}}$, chosen, where \citet{Hirata_2004} found that $N_{\mathrm{sub}}$ needs to be larger than $N_{\mathrm{bin}}^{3/2}$, where $N_{\mathrm{bin}}$ is the number of radial bins, and that the sub-box-length needs to be larger than the largest scale that is probed.
In this study, we are using 10 radial bins and a maximum scale of $20\mathrm{Mpc}/h$, which makes $N_{\mathrm{sub}}=64$ compatible with the above criterion.

The covariance is calculated, using all the jackknife realisations as follows:
\begin{align}
    C_{kl} &= \frac{N_{\mathrm{sub}}-1}{N_{\mathrm{sub}}} \sum_{n=1}^{N_{\mathrm{sub}}} (\psi^n_k - \bar{\psi_k})(\psi^n_l - \bar{\psi_l}) \\
    &\mathrm{with} \ \bar{\psi_k} = \frac{1}{N_{\mathrm{sub}}} \sum_{n=1}^{N_{\mathrm{sub}}}\psi_k^n\\
    &\mathrm{and} \ \psi \in [w_{g+},\tilde{\xi}_{g+,2}]
\end{align}
where $k,l$ run over all radial bins and $n$ over all jackknife realisations from $1$ to $N_{\mathrm{sub}}$.

\subsection{Selections}
\label{sec:md selections}

The position sample is defined by all galaxies with $M_\star>10^{8.35}\mathrm{M_\odot}/h$, which corresponds to $50$ stellar particles in the lowest resolution simulation (TNG300), and can therefore be considered the largest set of resolved galaxies with a consistent definition between simulations.
Due to boxsize differences, this selection corresponds to $407180$ galaxies in TNG300, $115208$ in Horizon-AGN and $22279$ in EAGLE, at $z=0$ (see Table \ref{table:num samples} for all sample sizes).
For the shape sample, we opt for a consistent stellar mass cut of $M_\star>10^{9.15}\mathrm{M_\odot}/h$, which corresponds to $300$ stellar particles in TNG300, and is therefore the largest consistent set of galaxies with converged shapes \citep{Chisari_2015}.
In subsequent sections, we use $\kappa_{\mathrm{rot}}$, \ $|v/\sigma|$, bulge-to-total ratio ($\mathrm{BTR}$) and $r-i$ colour to split the shape sample into disks and ellipticals (see Section \ref{sec:r distributions}).
The motivation for using these variables is based in the literature, as described in Section \ref{sec:intro}, where each of these variables has been used by one or multiple other works to make a split in disk and elliptical galaxies.
For consistency, many variables are remeasured from the particle data (in TNG300 and EAGLE) using a code developed for this reason, even if similar variables are available in simulation sub-halo catalogues.
The variables needed to reproduce the results in this paper are publicly available for TNG300\footnote{https://doi.org/10.5281/zenodo.18740718}, EAGLE\footnote{https://doi.org/10.5281/zenodo.18740922} and Horizon-AGN\footnote{https://doi.org/10.5281/zenodo.18740433}.

The rotational energy divided by the kinetic energy of the stellar particles, $\kappa_{\mathrm{rot}}$, can be used as a proxy to determine whether a galaxy is rotation or dispersion dominated. This is often used as a definition for disk (rotation dominated) and elliptical (dispersion dominated) galaxies.
$\kappa_{\mathrm{rot}}$ is measured in TNG300 and EAGLE and is defined as follows:
\begin{equation}
    \label{eq: kappa_rot}
    \kappa_{rot,s} = \frac{\sum^{N_s}_{n=1} \left[ \frac{1}{2} m_n (\hat{\vec{L}}_s\times\hat{\vec{r}}_n) \cdot \vec{v}_n \right]}
    {\sum^{N_s}_{n=1} \left[ \frac{1}{2} m_n \vec{v}_n \cdot \vec{v}_n \right]}
\end{equation}
where the $\hat{}$ -symbol denotes the direction of the vector and $\vec{v}_n$ is the velocity of each particle $n$ in galaxy $s$ relative to the centre of velocity of the galaxy, defined by:
\begin{align}
    \vec{v}_n &= \vec{\dot{x}}_n-\vec{\dot{x}}_s \label{eq: rel vel} \\
    \vec{\dot{x}}_s &= \frac{1}{M_s} \sum^{N_s}_{n=1} m_n \vec{\dot{x}}_n\\
\end{align}
with stellar particle velocity, $\vec{\dot{x}}_n$ and mass-weighted velocity of the stellar particles for each galaxy, $\vec{\dot{x}}_s$.
The angular momentum of the galaxy is given by:
\begin{align}
    \vec{L}_s &= \sum^{N_s}_{n=1} m_n \vec{r}_n \times \vec{v}_n \ .
    \label{eq: L}
\end{align}

$|v/\sigma|$ is the rotational velocity divided by the velocity dispersion of a galaxy, and therefore based on the same idea of defining disks and ellipticals by kinematics as $\kappa_{\mathrm{rot}}$.
It is measured following the procedure in \citet{Dubois_2016}.
First, a cylindrical coordinate system with the angular momentum (Eq. \ref{eq: L}) as $z$-axis and all coordinates ($R,\phi,z$) relative to the galaxy centre of mass and velocity ($\vec{r}_n$, $\vec{v}_n$) is adopted.
Second, the mass-weighted mean of the $\phi$ component of the velocity is computed, 
\begin{equation}
    \overline{v_{\phi,s}} = \frac{1}{M_s}\sum^{N_s}_{n=1} m_n v_{\phi,n} \ .
\end{equation}
Last, the velocity dispersion is computed as follows:
\begin{align}
    \overline{v_{k,s}} &= \frac{1}{M_s} \sum^{N_s}_{n=1} m_n v_{k,n} \\
    \overline{\Delta v_{k,s}} &= \frac{1}{M_s} \sum^{N_s}_{n=1} m_i (v_{k,n}-\overline{v_{k,s}})^2 \\
    \sigma &= \sqrt{\frac{\overline{\Delta v_{R,s}}+\overline{\Delta v_{\phi,s}}+\overline{\Delta v_{z,s}}}{3}}
\end{align}
where $k$ is $R,\phi$ or $z$ and $n$ runs over the particles of galaxy $s$ as before.
$|v/\sigma|$ is then defined as the absolute value of $\overline{v_{\phi,s}}/\sigma$.

The photometric colours used in this work are all from the available catalogues described in \citet{Nelson_2017} for TNG300, \citet{Trayford_2015} for EAGLE and \citet{Kaviraj_2017} for Horizon-AGN, and measured to be SDSS-like, mimicking the colour filters of the Sloan Digital Sky Survey (SDSS), including dust extinction.
Using colours to split a sample morphologically is one of the most widely used methods in simulations and observations.

\begin{figure*}
    \begin{center}
        \centering
    	\includegraphics[width=1.0\textwidth]{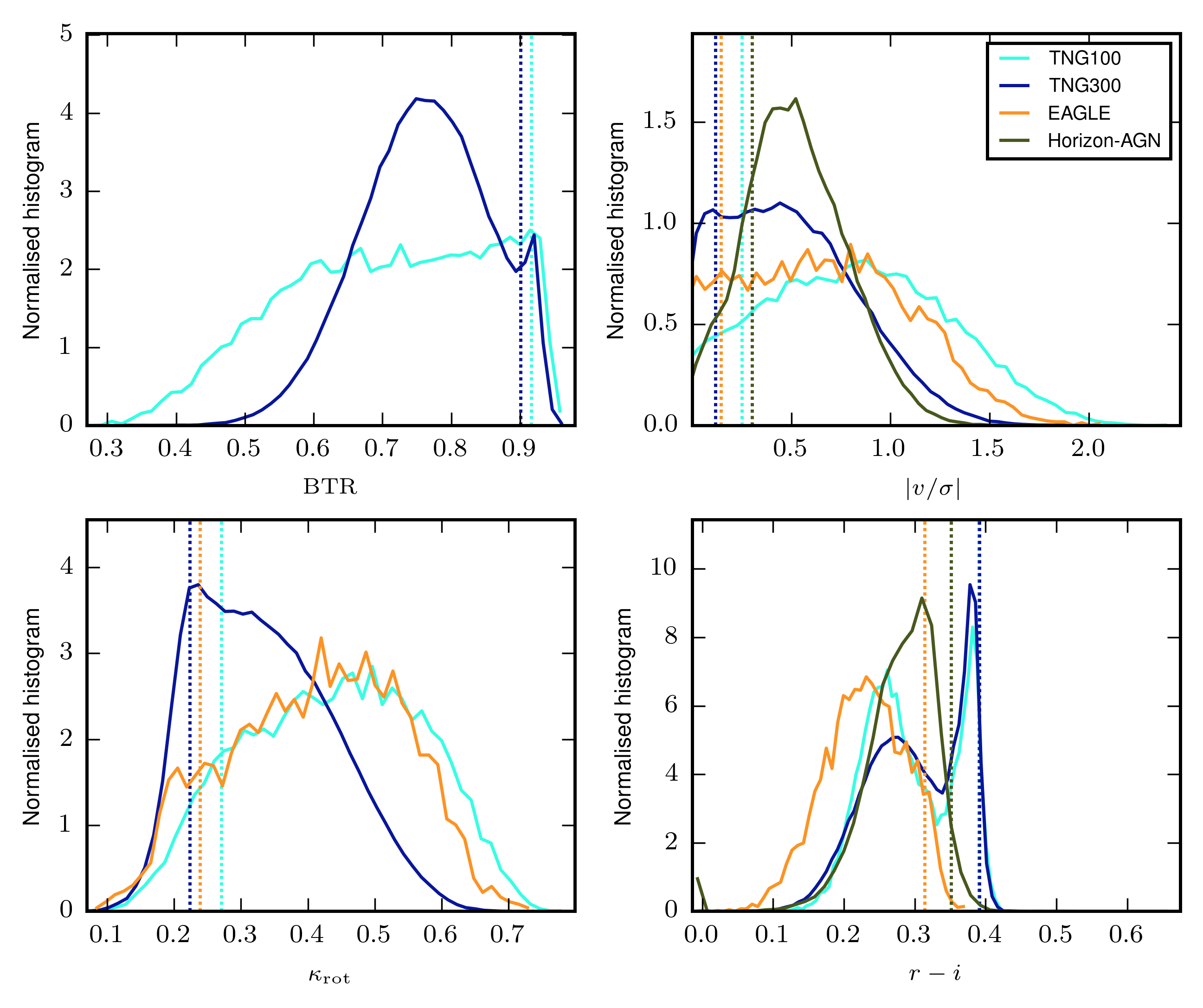}
    	\caption{Normalised distributions of $\mathrm{BTR}$ (top left), $|v/\sigma|$ (top right), $\kappa_{\mathrm{rot}}$ (bottom left) and $r-i$ (bottom right), which are used to split galaxies according to morphology, in TNG100 (light blue), TNG300 (dark blue), EAGLE (orange) and Horizon-AGN (green) at $z\approx0$.
        Vertical lines have been added to indicate the split between disks and elliptical galaxies.
        The variables show a range of distributions with varying features across simulations.
        The colour bi-modality in TNG simulations is not present in EAGLE and Horizon-AGN. Most likely this is because they produce many low-mass red galaxies, due to limited resolution, that are not included in this mass cut.}
        \label{fig:comp sim distr}
    \end{center}
\end{figure*}

The bulge-to-total ratio (BTR) is another variable based on the kinematic distinction between disks and ellipticals, by assigning stellar particles to the disk or bulge component of a galaxy based on their kinematic properties. 
This variable is not available for EAGLE and Horizon-AGN, and taken directly from the TNG300 catalogue \citep{Genel2015}, and therefore available for a limited number of galaxies, i.e. $M_\star>10^{9.3}\mathrm{M_\odot}/h$ at $z=0$ and $M_\star>10^{9.8}\mathrm{M_\odot}/h$ at higher redshifts.
Therefore, $\mathrm{BTR}$ is included only in Section \ref{sec:r distributions}, and then examined separately in Appendix \ref{app: BTR}.
The $\mathrm{BTR}$ defined as $1-$ the fractional mass of stars in a galaxy with $\epsilon>0.7$, and measured as follows.
First, the angular momentum of the stars within 10 times the stellar half-mass radius is computed.
Second, the system's z-axis is aligned with this angular momentum.
Then, the stellar particles taken into account are sorted by their binding energy.
For every stellar particle, the specific angular momentum, $J_z$ is calculated.
$J(E)$ is noted, the maximum angular momentum of the stellar particles between 50 before and 50 after the particle in question (in the sorted list).
Then, $\epsilon$ is computed for each particle by $\epsilon=J_z/J(E)$.
Finally, the bulge to total ratio is computed as defined above.

\section{Results}
\label{sec:results}

This section presents the results of the measurement and comparison of the IA correlations in TNG300, Horizon-AGN and EAGLE.
Section \ref{sec:r distributions} presents the distributions of the variables used to split the shape sample into disks and ellipticals and how they are used to select the samples, along with the stellar mass distributions of the samples.
Next, Sections \ref{sec:r AB} and \ref{sec:r DxE} present the IA correlations of disks and ellipticals around galaxies, and disks around ellipticals, respectively.
Section \ref{sec:r Disks} explores the correlations of disks around galaxies and ellipticals in Horizon-AGN further and Section \ref{sec:r Mass} delves into the relation between the correlations and their stellar mass distributions.

\subsection{Distributions}
\label{sec:r distributions}

Figure \ref{fig:comp sim distr} shows a comparison between the normalised distributions of the bulge-to-total ratio, $\mathrm{BTR}$ (top left panel), $|v/\sigma|$ (top right panel), $\kappa_{\mathrm{rot}}$ (bottom left panel) and the colour $r-i$ (bottom right panel), which are described in Section \ref{sec:md selections}, for the simulations TNG100 (light blue), TNG300 (dark blue), EAGLE (orange) and Horizon-AGN (green) at $z=0$ ($z=0.0556$ for Horizon-AGN), when available.
We use these variables in following sections to split the galaxy samples in disks and ellipticals, along the values added as horizontal dashed lines in the same colour.

\begin{table}[]
\caption{Number of galaxies in each sample for each simulation.}
    \centering
    \begin{tabular}{ccccc}
        Redshift & Simulation & Positions & Ellipticals  & Disks \\ \hline
         \multirow{3}{*}{0}  & TNG300 & 407180&20695 & 185880\\ 
         &Horizon-AGN & 115208 & 6568 & 59112 \\
         &EAGLE & 22279& 553&4972 \\ \hline
         \multirow{3}{*}{1} &  TNG300 & 407568&18291 & 164660\\ 
         &Horizon-AGN & 141546 & 6654 & 59887 \\
         &EAGLE &22897 &784 &7062 \\
    \end{tabular}
    \label{table:num samples}
\end{table}

\begin{figure*}
    \begin{center}
        \centering
    	\includegraphics[width=1.0\textwidth]{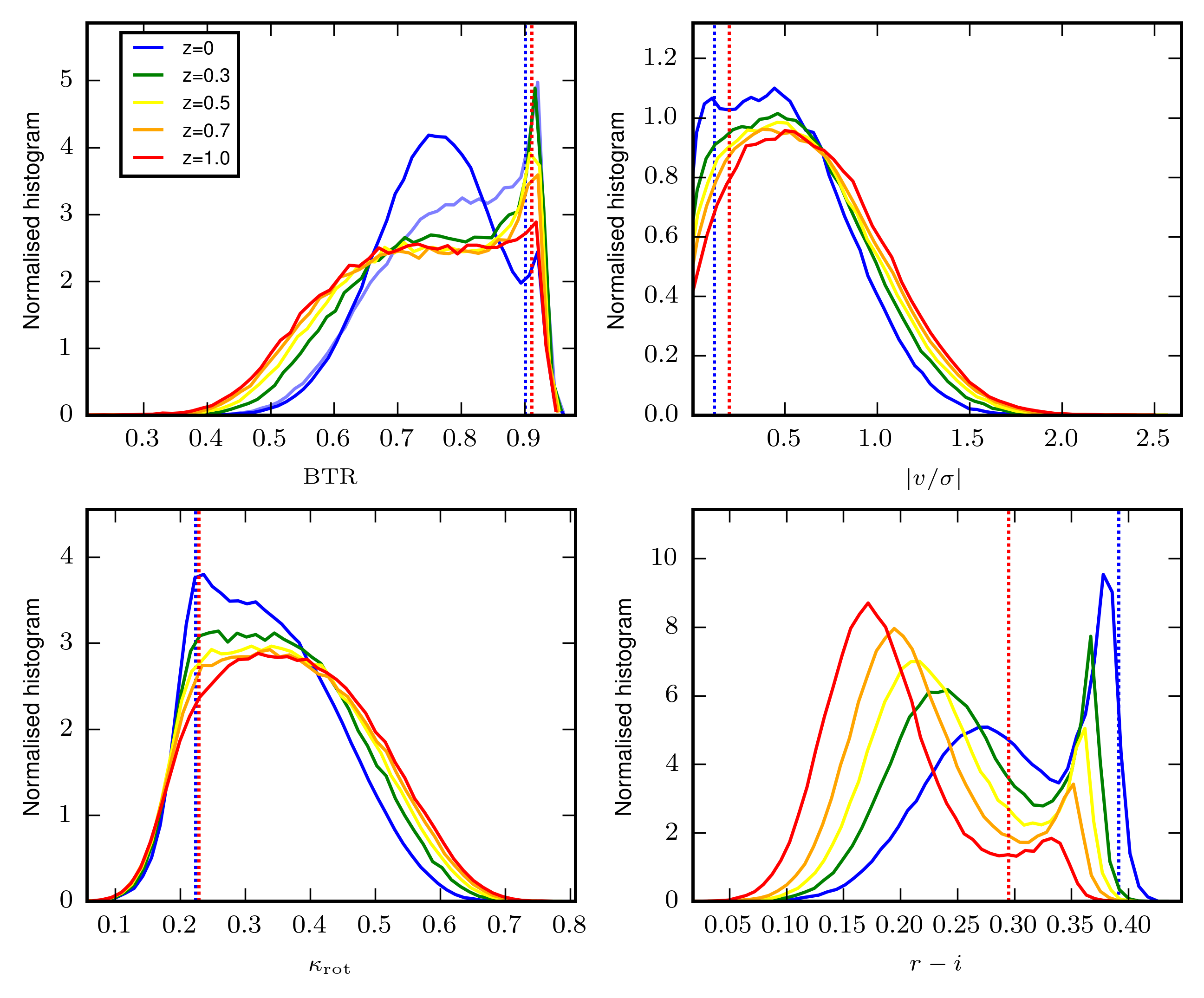}
    	\caption{Normalised distributions of $\mathrm{BTR}$ (top left), $|v/\sigma|$ (top right), $\kappa_{\mathrm{rot}}$ (bottom left) and $r-i$ (bottom right), which are used to split galaxies according to morphology, in TNG300 across a range of redshifts, $0\leq z\leq 1$.
        The $r-i$ distributions display a clear redshift evolution, whereas the other variables show only a mild evolution with redshift.}
        \label{fig:TNG300 distr z}
    \end{center}
\end{figure*}

The distributions of the variables that we use to perform the splits into disk and elliptical galaxy samples vary between simulations (Fig. \ref{fig:comp sim distr} - \ref{fig:Horizon-AGN distr z}).
Therefore, we opt for the abundance matching method to define our samples, i.e. using a set elliptical fraction of $10\%$ of the total sample of galaxies with well-resolved shapes, as simulations tend to produce less large ellipticals than found in observations \citep{Haslbauer_2022}.
In line with the consistent stellar mass cut for the samples across redshift, we also fix the elliptical fraction, although we know observations have found both parameters to evolve with redshift \citep{Mortlock_2013}.
Probing the redshift evolution of a consistent sample of galaxies, e.g. using merger trees, is outside the scope of this work.
The split value has been added as a vertical line for each simulation and variable in Figures \ \ref{fig:comp sim distr} - \ref{fig:Horizon-AGN distr z}. 
Table \ref{table:num samples} contains the number of galaxies in each sample, where the total shape sample is equal to disks + ellipticals and the selection using the variables in Fig. \ref{fig:comp sim distr} determines which galaxies make up each sample.

Figure \ref{fig:comp sim distr} shows that the simulations agree on the ranges of the variable values, sometimes by definition, but not on how the variables are distributed.
For TNG100 and TNG300, these differences can be ascribed to resolution differences that we explore in Appendix \ref{app:resolution}, together with a discussion on the influence of boxsize and resolution on the results.

The differences between TNG, EAGLE and Horizon-AGN cannot be as easily explained as the differences between simulations in the same suite.
Apart from varying resolutions and boxsizes, these simulations have different models and subgrid-physics recipies, as described in Section \ref{sec:md simulations}. 
Furthermore, where the kinematic parameters, $\kappa_{\mathrm{rot}}$ and $|v/\sigma|$, may be similar between TNG and EAGLE, the distributions of the colours have diverging features, of which only $r-i$ is shown in Figure \ref{fig:comp sim distr}.
TNG displays a bi-modality in its colours, as found in observational surveys, e.g. GAMA \citep{Taylor_2015} and SDSS \citep{Blanton_2003}, whereas this feature is not present in EAGLE or Horizon-AGN.
Colour bi-modality has been found in Horizon-AGN \citep{Kaviraj_2017}, albeit weaker than in observations.
As there are many low-mass red galaxies in Horizon-AGN \citep{Kaviraj_2017}, our mass cut is a likely cause of the uni-modality seen in Figure \ref{fig:comp sim distr}.
EAGLE also contains too many low-mass red galaxies due to limited numerical resolution.
For a detailed comparison between GAMA and EAGLE, see \citet{Trayford_2015}.

\subsubsection{Redshift evolution}
\label{sec: r d redshift evolution}

\begin{figure}
    	\centering
        \includegraphics[width=0.5\textwidth]   {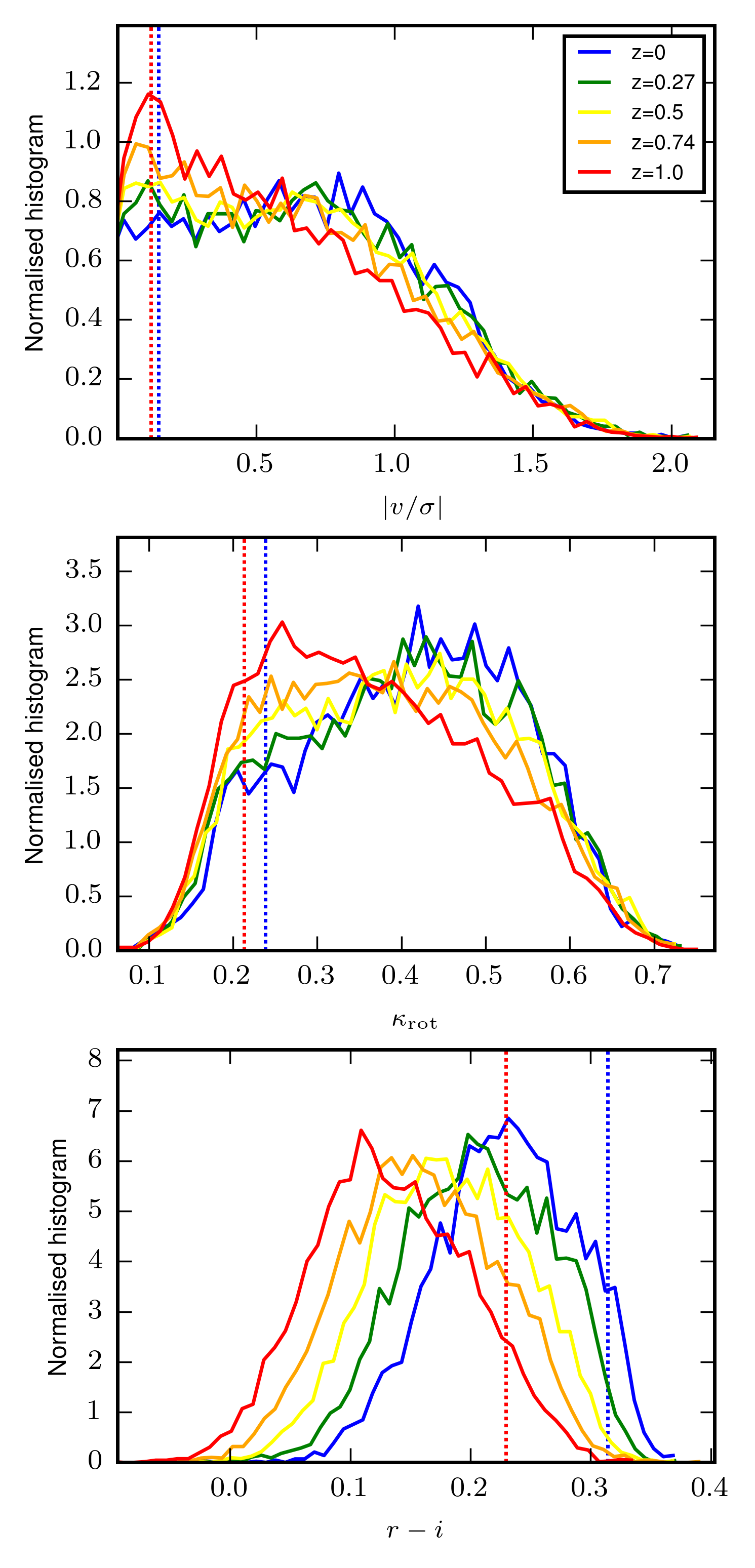}
    	\caption{Normalised distributions of $|v/\sigma|$ (top), $\kappa_{\mathrm{rot}}$ (middle) and $r-i$ (bottom), which are used to split galaxies according to morphology, in EAGLE across a range of redshifts, $0\leq z\leq 1$.
        All variables show a redshift evolution in their distributions. The kinematic variables $|v/\sigma|$ and $\kappa_{\mathrm{rot}}$ evolve opposite to expectations.}
        \label{fig:EAGLE distr z}
\end{figure}

\begin{figure}
    \centering
    \includegraphics[width=0.5\textwidth]{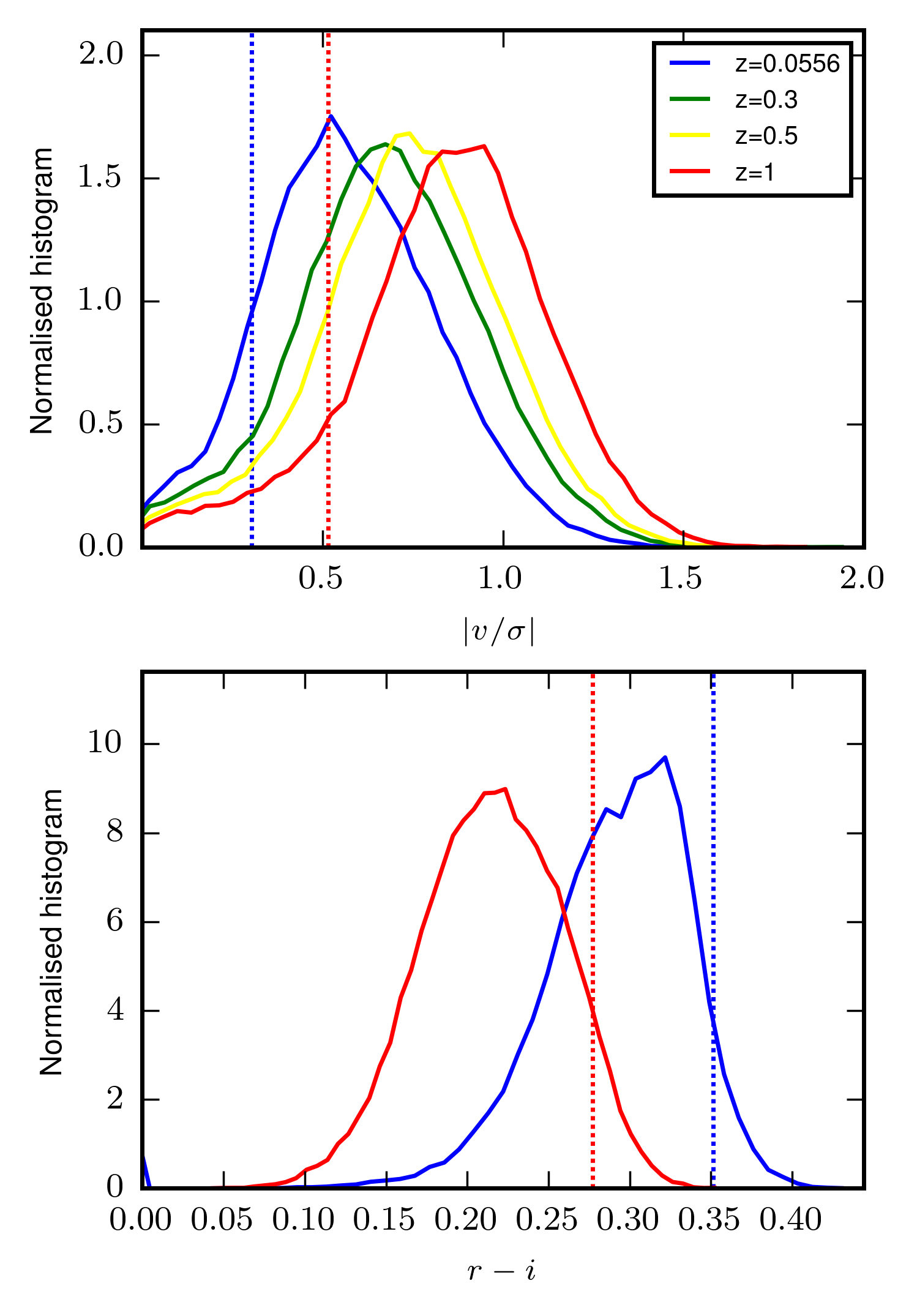}
    \caption{Normalised distributions of $|v/\sigma|$ (top) and $r-i$ (bottom), which are used to split galaxies according to morphology, in Horizon-AGN across a range of redshifts, $0\leq z\leq 1$.
    Both variables show redshift evolutions in line with expectations.}
    \label{fig:Horizon-AGN distr z}
\end{figure}

As an understanding of IA is not only crucial at $z=0$, but also at higher redshifts (see Section \ref{sec:intro}), it is important to consider the redshift evolution of the variables we intend to use to select our samples. 
Figures \ref{fig:TNG300 distr z}, \ref{fig:EAGLE distr z} and \ref{fig:Horizon-AGN distr z} show the variable distributions for each simulation between $0\leq z\leq 1$.
For the measurement of the intrinsic alignment correlations we show the results at $z=0$ and $z=1$, to probe the redshift evolution and because the redshift distribution of source galaxies in Stage IV surveys is estimated to peak at $z\sim1$.

Figure \ref{fig:TNG300 distr z} shows the redshift evolution for each of the variables in Figure \ref{fig:comp sim distr}, $\mathrm{BTR}$ (top left), $|v/\sigma|$ (top right), $\kappa_{\mathrm{rot}}$ (bottom left) and $r-i$ (bottom right), in TNG300.
For $z=0$ and $z=1$ the values used to split between disks and ellipticals are depicted in vertical dashed lines.
In agreement with observations \cite[e.g.][]{Maraston_2009}, the colours, represented by $r-i$, shift to lower (bluer) values with increasing redshift and the size of the peak on the red end of the colour spectrum diminishes.
This is due to the fact that the galaxy population is younger and more star-forming at higher redshift and there are less large red ellipticals, as they are formed later in the history of the universe.

The distributions of the kinematic variables, $\mathrm{BTR}$, \ $|v/\sigma|$ and $\kappa_{\mathrm{rot}}$, show a subtle redshift evolution in TNG300.
Recall from Section \ref{sec:md selections}, that the $\mathrm{BTR}$ catalogue is only available for $M_\star>10^{9.8}\mathrm{M_\odot}/h$ at $z>0$ and $M_\star>10^{9.3}\mathrm{M_\odot}/h$ at $z=0$, so the sample in this panel is different from those in the other panels in the figure.
The transparent blue line shows the distribution of $\mathrm{BTR}$ for $M_\star>10^{9.8}\mathrm{M_\odot}/h$ at $z=0$ for internal comparison and consistency.

The evolution of $\mathrm{BTR}$, $|v/\sigma|$, and $\kappa_{\mathrm{rot}}$ all point towards galaxies becoming more dispersion dominated, in stead of rotationally dominated as time passes, which is associated with an evolution from disk-like structures to ellipticals.
Note, however, that the sample of galaxies is not the same at each redshift, as the galaxies have not been tracked through time, but a consistent mass cut is chosen at each redshift instead.

We show these distributions to illustrate the evolution of our chosen samples, not to give an accurate representation of the redshift evolution of the properties of a galaxy population.
Figure \ref{fig:EAGLE distr z} shows the redshift evolution of the morphological variables in EAGLE, which include $|v/\sigma|$ (top panel), $\kappa_{\mathrm{rot}}$ (middle panel) and $r-i$ (bottom panel).
For $r-i$, the same shift towards bluer colours occurs with increasing redshift in EAGLE as in TNG300.

The kinematic variables, $|v/\sigma|$ and $\kappa_{\mathrm{rot}}$, show a clear redshift evolution in EAGLE.
Contrary to TNG300, the $|v/\sigma|$ distribution skews towards lower values as we go back in time (higher redshift), indicating that there are more galaxies with a higher velocity dispersion, when compared to their rotational velocity.
Therefore, Figure \ref{fig:EAGLE distr z} shows an evolution from elliptical galaxies at high redshift, to disk galaxies at low redshift.
The $\kappa_{\mathrm{rot}}$ distribution evolves in a similar way: the peak skews towards lower values of $\kappa_{\mathrm{rot}}$ at higher redshift, which are generally associated with elliptical galaxies.
Why these two simulations show opposing behaviour is unclear and left to future research.

\begin{figure*}
    \begin{center}
        \centering
        \includegraphics[width=1.0\textwidth]{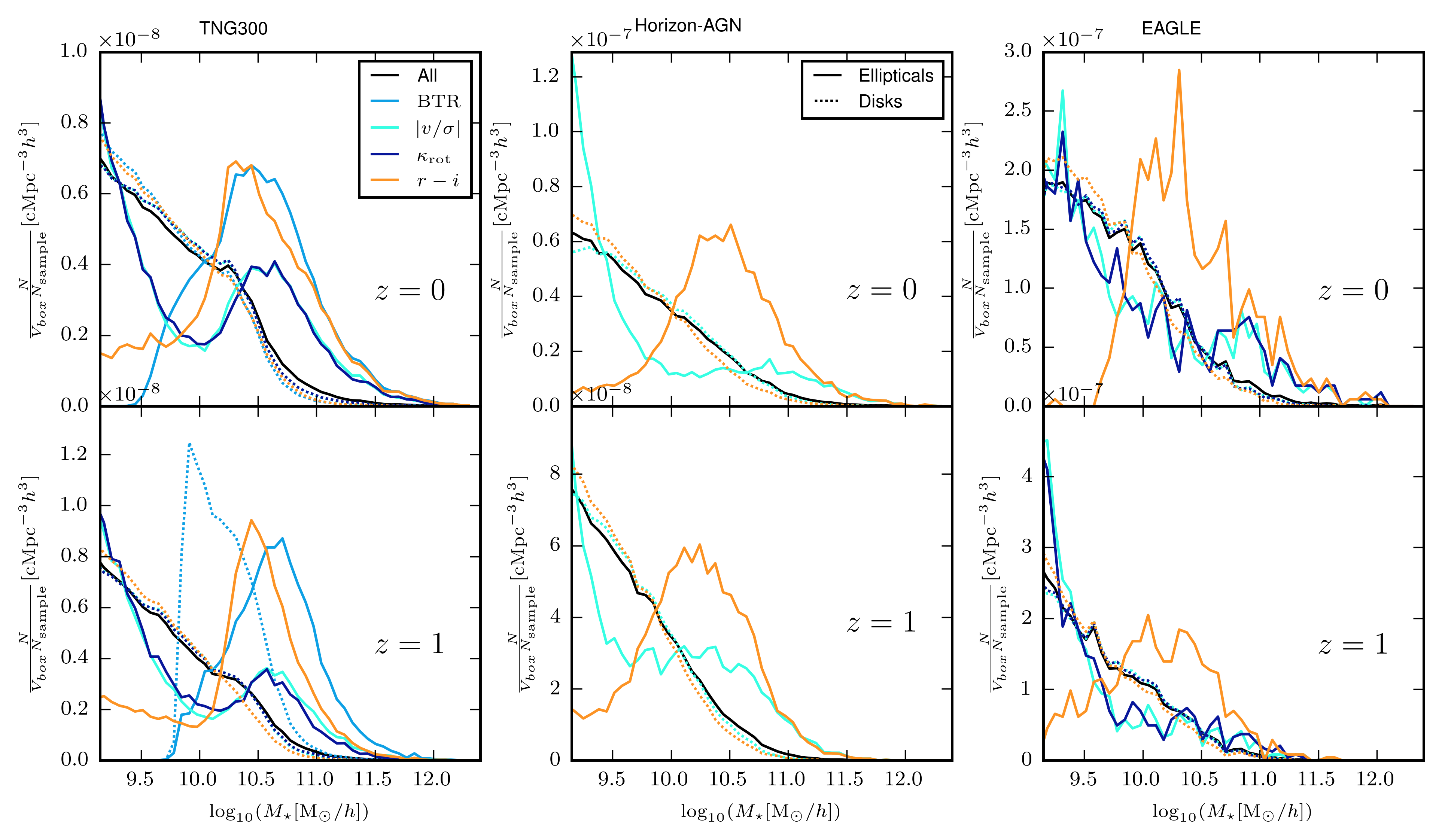}
        \caption{Galaxy stellar mass distributions at $z=0$ (top) and $z=1$ (bottom) of the full galaxy sample (black) and elliptical (continuous line style) and disk (dashed) samples, defined based on a selection in $\mathrm{BTR}$ (medium blue), $|v/\sigma|$ (light blue), $\kappa_{\mathrm{rot}}$ (dark blue) or $r-i$ (orange), in TNG300 (left), Horizon-AGN (middle) and EAGLE (right).
        The distributions are normalised by simulation volume and sample size, where the latter can be found in table \ref{table:num samples}.
        The choice of variable used in the selection of the elliptical sample has a large impact on the resulting galaxy stellar mass distribution of the sample.
        The stellar mass distributions of the selections evolve with redshift.}
        \label{fig:Mass distr}
    \end{center}
\end{figure*}

Figure \ref{fig:Horizon-AGN distr z} shows the redshift evolution of the distributions of $|v/\sigma|$ (top panel) and the colour $r-i$ (bottom panel) for Horizon-AGN.
For $r-i$ (bottom panel, only available to us at $=0.0556$ and $z=1$), we see the same general trend as in EAGLE and TNG300: the peak shifts towards bluer colours with increasing redshift.
The $|v/\sigma|$ distributions (top panel) show a clear redshift evolution towards higher $|v/\sigma|$ values with increasing redshift, indicating a higher abundance of disk galaxies at higher redshifts.
This evolution is the exact opposite of that in EAGLE.

\subsubsection{Mass distributions}
\label{sec:r d defining de}

Figure \ref{fig:Mass distr} shows the normalised galaxy stellar mass distributions of the full sample of galaxies with well-resolved shapes (black lines) and the samples of ellipticals (continuous lines) and disks (dashed lines), when defined using $\mathrm{BTR}$ (medium blue), $|v/\sigma|$ (light blue), $\kappa_{\mathrm{rot}}$ (dark blue) or $r-i$ (orange), at $z=0$ (top) and $z=1$ (bottom) in TNG300 (left), Horizon-AGN (middle) and EAGLE (right).
While there are some discrepancies between the simulations, the stellar mass distributions of the full samples agree well.
The mass distributions of the disk samples of every morphological definition match the full sample distributions closely in all shown cases, as they contain $90\%$ of the galaxies of the full sample.

For the elliptical samples at $z=0$, the galaxy stellar mass distributions vary significantly when comparing selections based on different variables within one simulation and are similar when comparing the elliptical samples defined by the same variable across simulations.
Within TNG300 (top left panel), selecting the galaxies with the highest $10\%$ $\mathrm{BTR}$ (medium blue) or $r-i$ (orange), results in a similar mass distribution, that favours higher mass galaxies and is peaked at $\mathrm{log}(M_\star[h/M_\odot])\sim10.5$.
In comparison, the elliptical samples selected based on $\kappa_{\mathrm{rot}}$ (dark blue) or $|v/\sigma|$ (light blue) include more low-mass galaxies ($\mathrm{log}(M_\star[h/M_\odot])\lesssim10$), resulting in a bimodal distribution with its high-mass peak also at $\mathrm{log}(M_\star[h/M_\odot])\sim10.5$.
In EAGLE (top right panel) and Horizon-AGN, the selection of ellipticals based on $r-i$, $\kappa_{\mathrm{rot}}$ and $|v/\sigma|$ have similar mass distributions as in TNG300, although in EAGLE the distributions are noisier due to the smaller sample size.
Based on these galaxy stellar mass distributions, we can discern two groups: $|v/\sigma|$ \& $\kappa_{\mathrm{rot}}$, which include a large fraction of low-mass galaxies; and $\mathrm{BTR}$ \& $r-i$, which do not.

\begin{figure*}
    \begin{center}
        \centering
    	\includegraphics[width=1.0\textwidth]{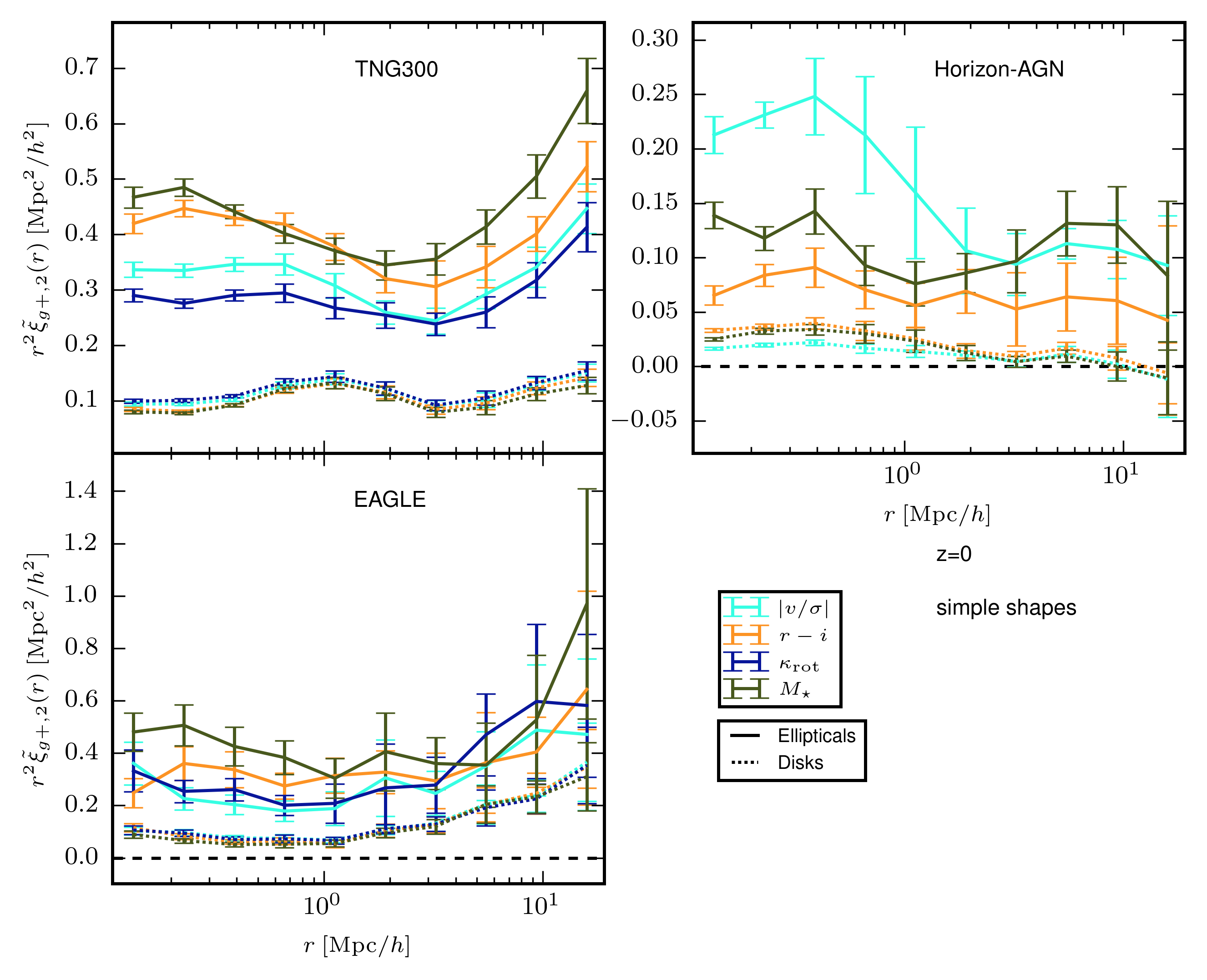}
    	\caption{
        The quadrupoles of ellipticals (continuous lines) and disks (dashed) around all resolved galaxies at $z=0$ in TNG300 (top left), Horizon-AGN (top right) and EAGLE (bottom left).
        The shape samples are defined by the values of the $|v/\sigma|$ (light blue), $\kappa_{\mathrm{rot}}$ (dark blue), $r-i$ (orange) or $M_\star$ (green) of the galaxies.
        The shapes are calculated using the simple inertia tensor.
        The simulations agree on the scale dependence and sign of the signal. 
        Furthermore, the amplitudes of the elliptical correlations are all higher than those of the disks.
        The relative amplitudes of the morphological definitions differ between simulations.
        }
        \label{fig:AB s0}
    \end{center}
\end{figure*}

At $z=1$, the stellar mass distributions, shown in the lower panels of Figure \ref{fig:Mass distr}, are similar to the distributions at $z=0$, with the most notable change being that there are less high mass galaxies in the samples.
Note that the disk sample of $\mathrm{BTR}$ (dashed blue line, bottom left panel) behaves differently due to the implicit mass cut at $\mathrm{log}(M_\star[h/M_\odot])\sim9.8$, but agrees with the full sample when the mass cut is chosen to be consistent with the $\mathrm{BTR}$ catalogue (not shown here).

\subsection{Disks and ellipticals around galaxies}
\label{sec:r AB}

In this section, we present the IA quadrupoles of disk and elliptical shape samples, as defined previously, around the positions of all resolved galaxies for TNG300, EAGLE and Horizon-AGN.
Furthermore, we explore the differences between using simple and reduced shapes at two redshifts, $z=0$ and $z=1$, in TNG300. 

Figure \ref{fig:AB s0} shows the quadrupoles for disks (dashed lines) and ellipticals (continuous lines) around all resolved galaxies for TNG300 (top left panel), Horizon-AGN (top right panel) and EAGLE (bottom left panel) at $z=0$.
The galaxy shapes are defined by the simple inertia tensor as described in Section \ref{sec:md shapes}.
The shape sample is split into disks and ellipticals as described previously, according to $r-i$ colour (orange), $|v/\sigma|$ (light blue), the $\kappa_{\mathrm{rot}}$ (dark blue) or stellar mass ($M_\star$, dark green), where stellar mass is not a true morphological split, but has been added as a reference.

\begin{figure*}
    \begin{center}
        \centering
    	\includegraphics[width=1.0\textwidth]{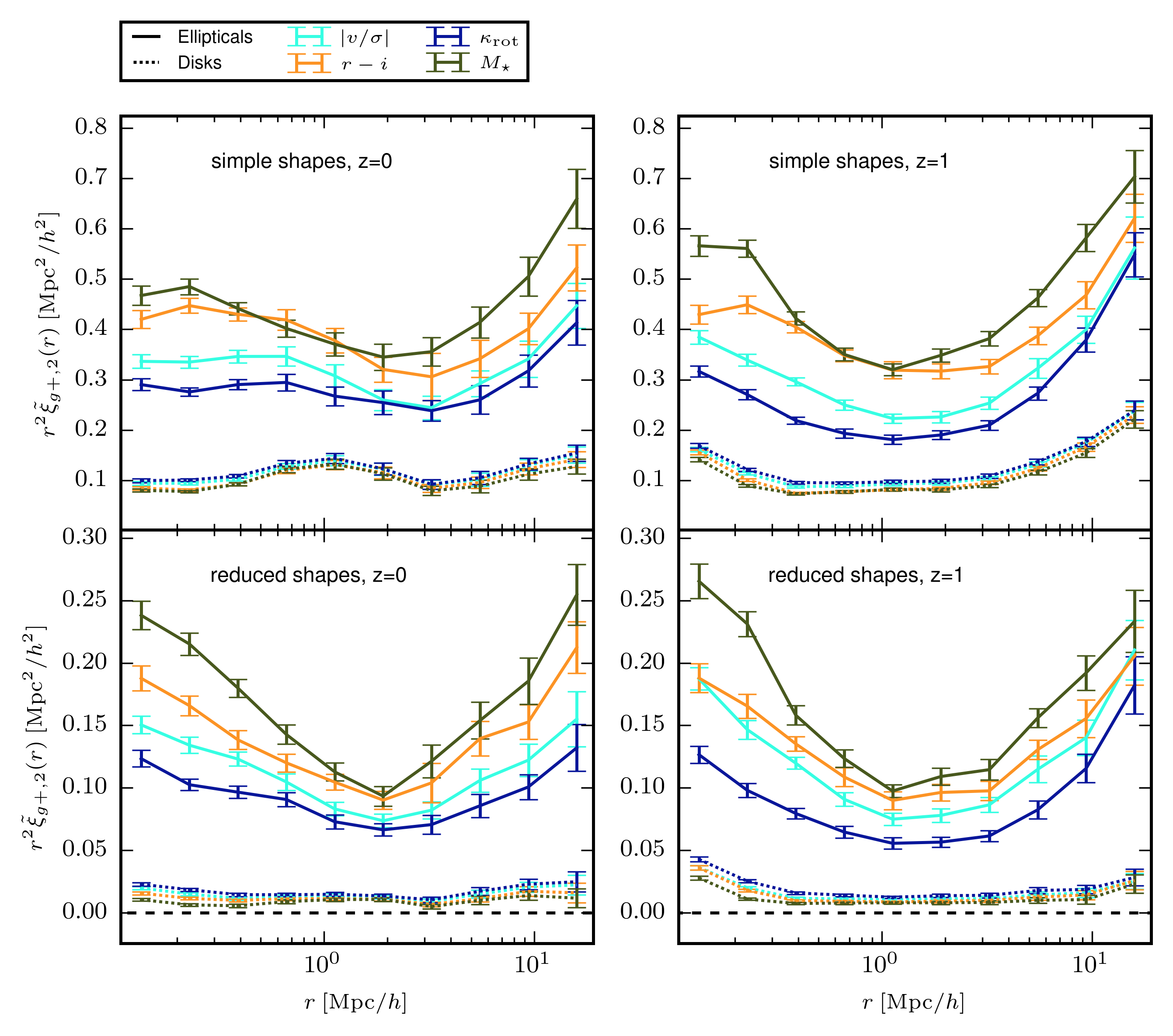}
    	\caption{
        The quadrupoles of ellipticals (continuous lines) and disks (dashed) around all resolved galaxies in TNG300 at $z=0$ (left) and $z=1$ (right) for shapes calculated using the simple (top) or reduced (bottom) inertia tensor.
        The shape samples are defined by the values of the $|v/\sigma|$ (light blue), $\kappa_{\mathrm{rot}}$ (dark blue), $r-i$ (orange) or $M_\star$ (green) of the galaxies.
        The scale dependence and ordering of the amplitudes of the different samples persist across redshift and shape definition indicating an underlying robustness of the signals.
        }
        \label{fig:AB TNG300}
    \end{center}
\end{figure*}

All simulations agree on the main trends in the quadrupoles of both the disks and ellipticals around galaxies, as shown in Figure \ref{fig:AB s0}.
For each simulation and selection made, the signals are all positive or consistent with null.
Moreover, the simulations agree that the amplitude of the quadrupoles of the ellipticals is higher than that of the disks.
Furthermore, multiplying $\tilde{\xi}_{g+,2}(r)$ by $r^2$ removes most of the scale dependence in TNG, EAGLE.
In Horizon-AGN $r^2\tilde{\xi}_{g+,2}(r)$ is also flat, with the exception of the ellipticals around galaxies correlation defined by $|v/\sigma|$ (light blue), which shows a higher amplitude at small scales, i.e. $r\lesssim1\mathrm{Mpc}/h$. 
The variation in amplitude based on morphological definition within one simulation is similar to the range between simulations.

The variables used to define the disk and elliptical galaxy samples are all strongly correlated with the stellar mass of the galaxies.
As we know from previous works \cite[e.g.][]{Chisari_2015,Tenneti_2016,Samuroff_2021}, stellar mass also has a strong correlation to the signal strength of the galaxy alignment correlation functions. 
The morphological definition used to select ellipticals leads to different signal amplitudes in TNG300, that can largely be traced back to their underlying stellar mass distributions (Fig. \ref{fig:Mass distr}), as also supported by the amplitudes of the stellar mass split (green).
At $r\gtrsim2\mathrm{Mpc}/h$, the IA signal amplitude follows the trend where more high-mass galaxies in the sample corresponds to a higher amplitude.
At $r\lesssim2\mathrm{Mpc}/h$, this is also largely the case, as the amplitude of the $r-i$ elliptical signal is higher than those of $|v/\sigma|$ and $\kappa_{\mathrm{rot}}$.
However, at these smaller scales, we are inside the haloes and sub-grid physics models play a larger role, which is why the $r-i$ elliptical quadrupoles overlap with the $M_\star$ ones despite their diverging mass distributions; and the $|v/\sigma|$ and $\kappa_{\mathrm{rot}}$ quadrupoles are distinct, despite their overlapping mass distributions.
Therefore, at these non-linear scales, we can conclude that not everything is determined by mass.
This is further explored in Section \ref{sec:r Mass}.

In EAGLE, due to the size of the error bars (number of galaxies), no robust conclusions can be drawn about the differences between morphological definitions.

In Horizon-AGN (top right panel of Figure \ref{fig:AB s0}) we observe a different trend. 
The differences in amplitude between elliptical samples do not correlate with the underlying stellar mass distributions in the same way as in TNG300.
Considering Figure \ref{fig:Mass distr}, we would expect the quadrupole with the shape sample defined by $|v/\sigma|$ to have a lower amplitude than the one defined by $r-i$, as the $r-i$ sample includes much more high-mass galaxies.
Figure \ref{fig:AB s0} shows the opposite.
This means that either the underlying mass distributions are not dictating the signals, or the IA amplitude does not monotonically increase with mass in Horizon-AGN.

To infer how the selections in morphological variables and simulations influence the correlation functions, we explore their robustness under an alternate definition of galaxy shapes and redshift evolution.
Figure \ref{fig:AB TNG300} shows the quadrupoles ($\times r^2$) of disks (dashed lines) and ellipticals (continuous lines) around all galaxies (defined as in Figure \ref{fig:AB s0}) for TNG300 for simple (top panels) and reduced (bottom panels) shapes at $z=0$ (left panels) and $z=1$ (right panels).
The colours correspond to the different variables being used to define the elliptical and disk sample, as described previously.

Figure \ref{fig:AB TNG300} shows a robustness in the main trends of the signals across both shape definition and redshift evolution.
As in Figure \ref{fig:AB s0}, the split in stellar mass has been added to the plots in Figure \ref{fig:AB TNG300} and again, the mass samples follow the main trends of the corresponding variable samples.
The amplitudes of the signals are similar across redshift, albeit somewhat diminished when going from simple shapes to reduced shapes.
This is to be expected as reduced shapes are generally rounder, due to the down-weighting of the outer regions of the galaxies, and therefore lower the correlation signal amplitude.
Furthermore, the correlations using the reduced shapes at $z=1$ show an extra scale dependence, when compared to the case of simple shapes at $z=0$.

Comparing the ellipticals around galaxies correlations between top and the bottom left panel in Figure \ref{fig:AB TNG300}, i.e. shapes measured by the simple versus the reduced inertia tensor Eq. \ref{eq: reduced inertia tensor}, 
we see that the most notable difference is the distinction between the high stellar mass sample (dark green) and the $r-i$ elliptical sample (orange), at small scales ($r\lesssim1\mathrm{Mpc}/h$).
These two signals were not distinct for simple shapes, but for the reduced shapes the amplitude of the high stellar mass sample is clearly higher, indicating a possible difference between the influence of mass and colour on the IA in inner and outer regions of galaxies.

\begin{figure*}
    \begin{center}
        \centering
    	\includegraphics[width=1.0\textwidth]{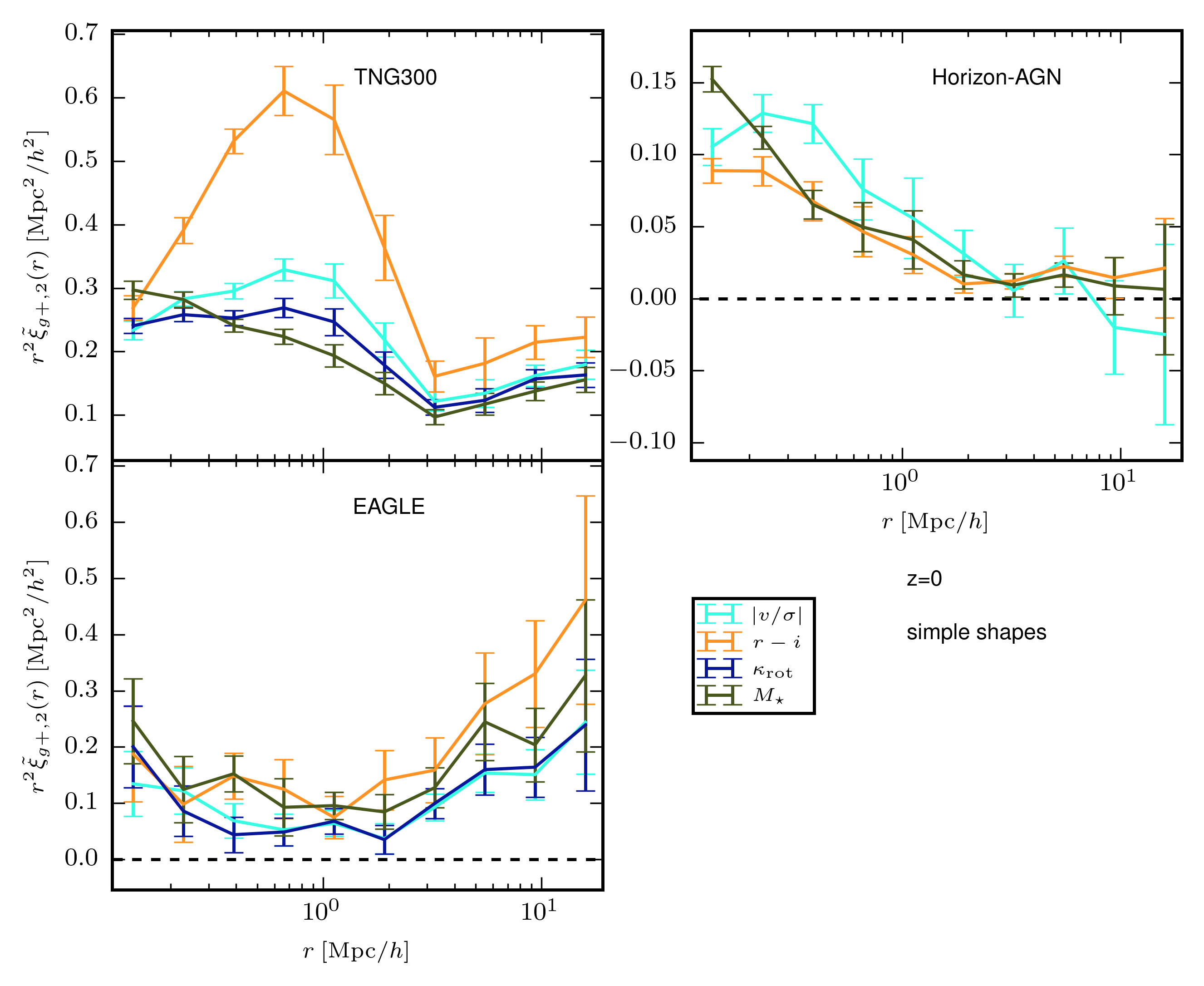}
    	\caption{
        The quadrupoles of disks around ellipticals at $z=0$ in TNG300 (top left), Horizon-AGN (top right) and EAGLE (bottom left).
        The disk and elliptical samples are defined by the values of the $|v/\sigma|$ (light blue), $\kappa_{\mathrm{rot}}$ (dark blue), $r-i$ (orange) or $M_\star$ (green) of the galaxies.
        The shapes are calculated using the simple inertia tensor.
        For this type of alignment, the simulations show differences in secondary scale-dependence and impact of morphological definition.
        In particular, the amplitude difference in TNG300 between the $r-i$ and other signals at intermediate scales indicates a fundamental difference between the two selections.
        }
        \label{fig:DxE s0}
    \end{center}
\end{figure*}

Comparing the reduced shape measurements at $z=0$ (bottom left panel) and $z=1$ (bottom right panel), reveals that the ellipticals defined by their $|v/\sigma|$ increase their IA signal amplitude at $z=1$, despite having relatively fewer high mass galaxies in their samples. 
Despite their vastly different underlying stellar mass distributions (see Figure \ref{fig:AB TNG300}), the quadrupoles measured for the ellipticals when defined by $|v/\sigma|$ or $r-i$ agree quite well at $z=1$ for reduced shapes, whereas the amplitude of the $\kappa_{\mathrm{rot}}$ signal is much lower, despite the sample having the same underlying stellar mass distribution as $|v/\sigma|$, again advocating for the influence of sub-grid physics over stellar mass.

Note that for Figure \ref{fig:AB TNG300} the same can be said for the disk signals as for Figure \ref{fig:AB s0}: the groupings found in the ellipticals are also present in the disk signals, albeit with opposite relative amplitudes.
As the full shape sample (disks + ellipticals) is always the same for every split, this reflection is to be expected.

Figures similar to Figure \ref{fig:AB TNG300} have been made for EAGLE and Horizon-AGN, and are presented in Appendix \ref{app:AB EAGLE HorizonAGN}.
For both EAGLE and Horizon-AGN, the signal shape and sign is robust across redshift and shape measurement method and the amplitude for reduced shapes is lower than for simple shapes.

\subsection{Disks around ellipticals}
\label{sec:r DxE} 

All measured IA correlations of disks (and ellipticals) around galaxies measured in Section \ref{sec:r AB} are positive or null, and therefore in disagreement with the negative disk signal measured in previous works \citep{Chisari_2016,Kraljic_2020}.
This section presents the IA correlation of the shapes of disks around the positions of ellipticals for all chosen morphological definitions in the simulations, matching the choice used in \citet{Chisari_2016}.
In this section, we use the same definitions of disks and ellipticals as in Section \ref{sec:r AB}, which is different to the one used in \citet{Chisari_2016}.

\begin{figure*}
    \begin{center}
        \centering
    	\includegraphics[width=1.0\textwidth]{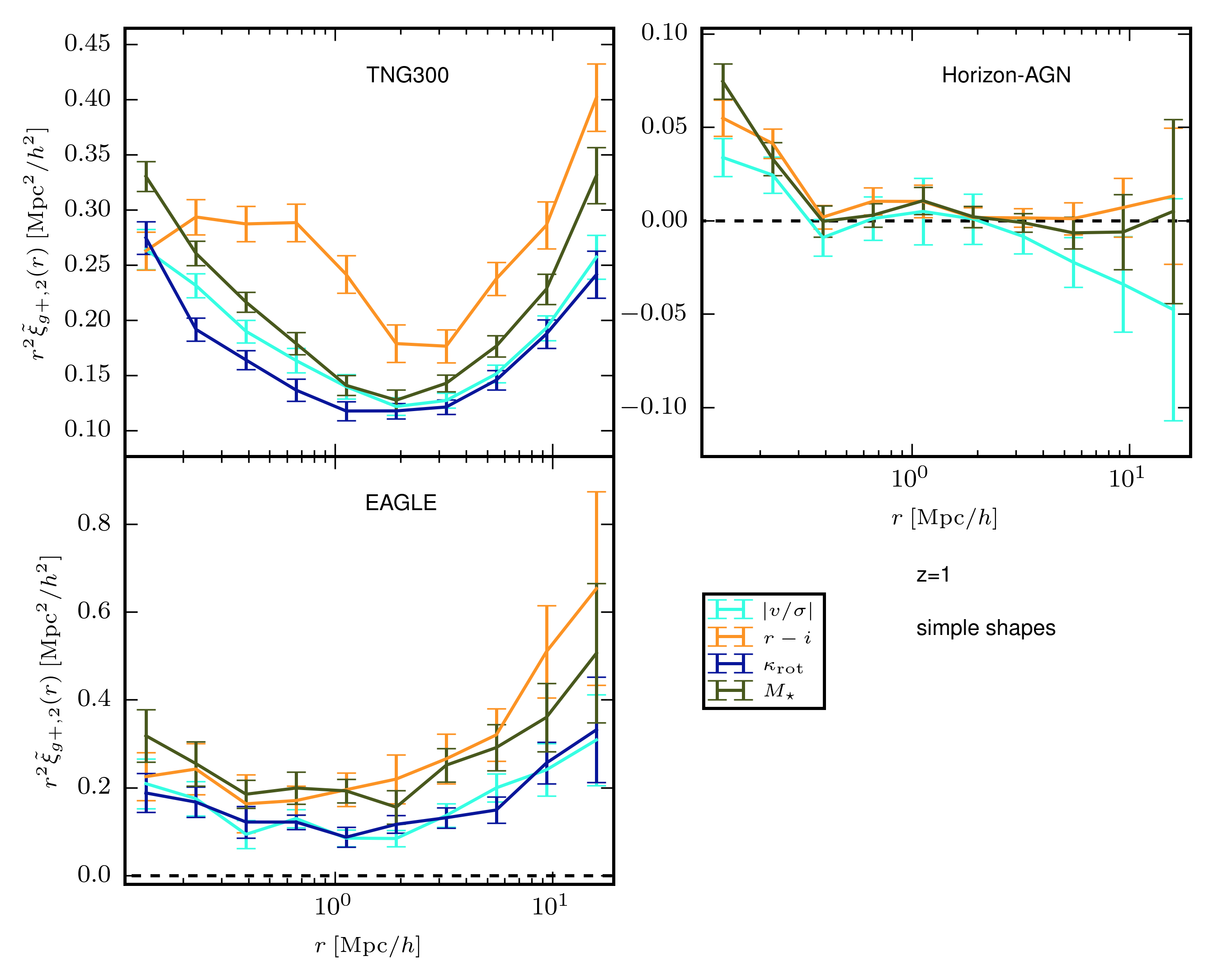}
    	\caption{
        The quadrupoles of disks around ellipticals at $z=1$ in TNG300 (top left), Horizon-AGN (top right) and EAGLE (bottom left).
        The disk and elliptical samples are defined by the values of the $|v/\sigma|$ (light blue), $\kappa_{\mathrm{rot}}$ (dark blue), $r-i$ (orange) or $M_\star$ (green) of the galaxies.
        The shapes are calculated using the simple inertia tensor.
        The amplitude of Horizon-AGN diminishes with redshift, EAGLE is robust and for TNG300 there is some redshift evolution of all correlations.
        }
        \label{fig:DxE s1}
    \end{center}
\end{figure*}

Figure \ref{fig:DxE s0} shows the quadrupoles ($\times r^2$) for the shapes of disks around the positions of ellipticals, in TNG300 (top left panel), Horizon-AGN (top right) and EAGLE (bottom left) at $z=0$.
The shapes are measured using the simple inertia tensor.
Similar to the signals of disks and ellipticals around galaxies (Figure \ref{fig:AB s0}), the simulations agree with each other on the main trends of the disks around ellipticals correlations in Figure \ref{fig:DxE s0}:
the main $r$-dependence of the quadrupoles is quadratic and the signals are positive.
While $r^2\tilde{\xi}_{g+}$ does not have a strong scale dependence in EAGLE and for most TNG300 morphological definitions, there is a residual scale dependence for the disks selected by $r-i$ around ellipticals correlation in TNG300 and for all morphological definitions in Horizon-AGN.
The $r-i$ correlation in TNG300 shows a peak is at $r\sim 1$ and the Horizon-AGN $r^2\tilde{\xi}_{g+}$ measurements decrease with $r$.
For EAGLE (bottom left panel) there is no distinction between the choices in morphological definition as all the signals agree within error bars.
In Horizon-AGN (top right panel), there is no clear distinction between morphological definitions either, contrary to the ellipticals around galaxies signal shown in Fig. \ref{fig:AB s0}.

While the morphological definitions have little effect on the correlation between disk shapes and elliptical positions in the 2-halo regime in TNG300 (top left panel), within the halos ($r<2\mathrm{Mpc}/h$), the correlations are distinct.
The correlations defined by $|v/\sigma|$ and $\kappa_{\mathrm{rot}}$ agree well, but they are distinct from the $r-i$ correlation.
As the alignment amplitude of blue around red galaxies is $\sim2.5$ times as large (at $r\approx0.7\mathrm{Mpc}/h$) as that of low around high mass galaxies, which agrees well with the kinematic variables, the offset of the $r-i$ correlation cannot be fully explained by the underlying stellar mass distributions.
Galaxies selected by colour show a different alignment behaviour than those selected by their dynamics or their mass in TNG300.
While both types of signals (disks and ellipticals around galaxies and disks around ellipticals) show signs of influence by sub-grid physics, the divergence of the disks around elliptical alignment when selected by colour provides complementary insights into the mechanisms behind IA.

\begin{figure*}
    \begin{center}
        \centering
    	\includegraphics[width=1.0\textwidth]{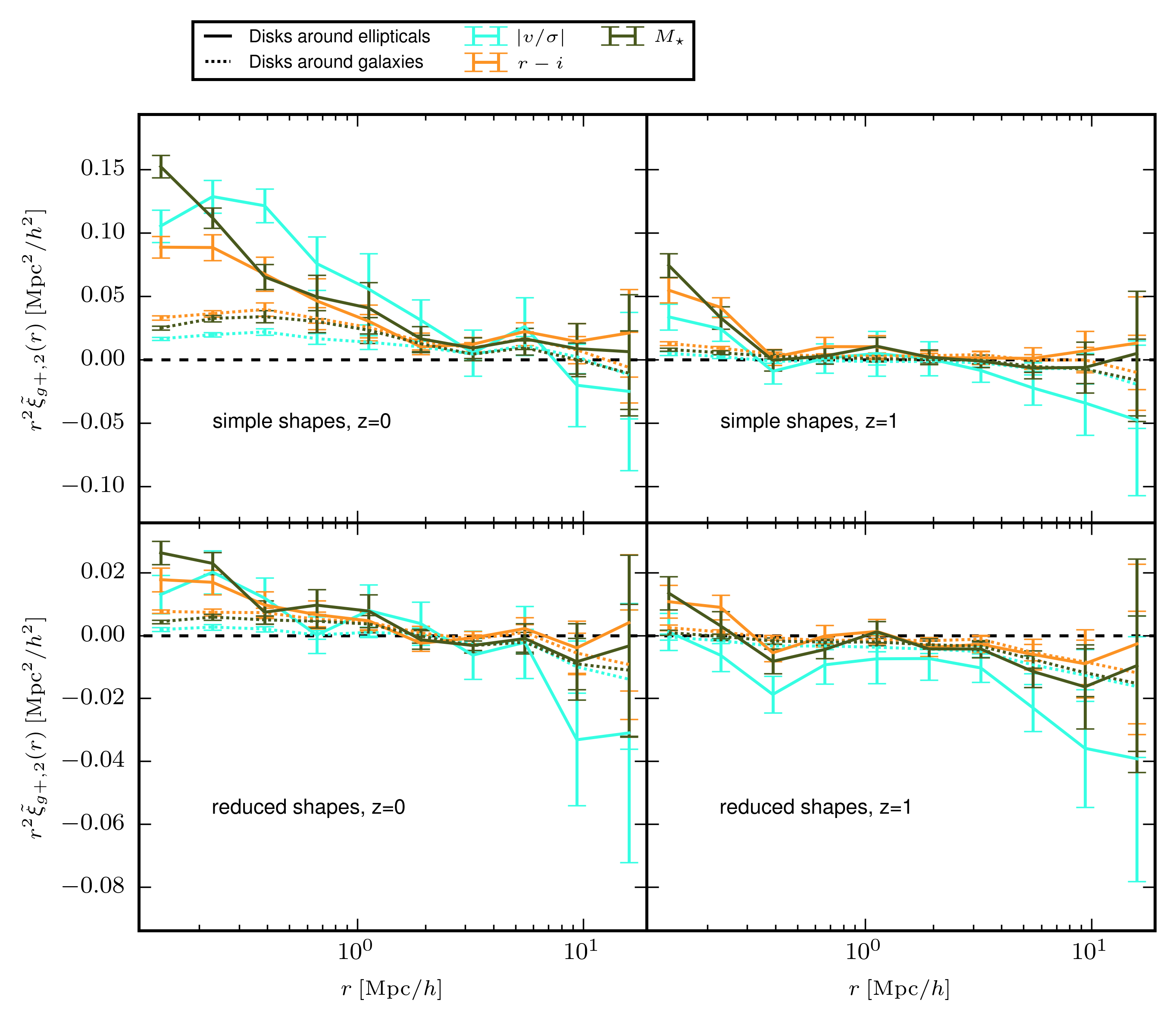}
    	\caption{
        The quadrupoles of disks around ellipticals (continuous lines) and disks around all resolved galaxies (dashed) in Horizon-AGN at $z=0$ (left) and $z=1$ (right) for shapes calculated using the simple (top) or reduced (bottom) inertia tensor.
        The shape samples are defined by the values of the $|v/\sigma|$ (light blue), $r-i$ (orange) or $M_\star$ (green) of the galaxies.
        The Horizon-AGN disk signals diminish with redshift and for reduced shapes, compared to simple shapes. The disks around ellipticals signal defined by $|v/\sigma|$ for reduced shapes at $z=1$ is negative.
        }
        \label{fig:AB DxE HorizonAGN}
    \end{center}
\end{figure*}

Figure \ref{fig:DxE s1} displays the quadrupoles of disks around ellipticals as presented in Figure \ref{fig:DxE s0} at $z=1$.
All colours are the same as previous figures, and again we use the simple inertia tensor to measure the galaxy shapes.
Note that the mass cuts used in the selection of galaxies remain constant with redshift and therefore, Figure \ref{fig:DxE s1} represents the alignments of a different population of galaxies compared to Figure \ref{fig:DxE s0}.
While the measurements of $r^2\tilde{\xi}_{g+,2}$ in EAGLE do not show a strong scale dependence at $z=1$, similar to $z=0$, TNG300 has a scale dependence that decreases with $r$ in the 1-halo regime and increases with $r$ in the 2-halo regime.
Like TNG300, Horizon-AGN has a diminished amplitude at $z=1$, which results in flat $r^2\tilde{\xi}_{g+,2}$ measurements, consistent with null at most scales.

Comparing the top left panels (TNG300) of Figures \ref{fig:DxE s1} and \ref{fig:DxE s0}, reveals a reduced offset of the $r-i$ (orange) disks around ellipticals correlation to the $M_\star$ correlation (green) at $r\approx0.7\mathrm{Mpc}/h$, from $\sim2.5$ at $z=0$ to $\sim1.7$ at $z=1$.

In Horizon-AGN, the $|v/\sigma|$ disks around ellipticals correlation (light blue) starts to reveal a negative at large scales, $r\gtrsim5\mathrm{Mpc}/h$.
To fully uncover for which types of measurements the correlation is below zero in Horizon-AGN, we explore both disk around ellipticals and disks around all galaxies correlations for simple and reduced shapes at $z=0$ and $z=1$ in the following section.

\subsection{Disks in Horizon-AGN}
\label{sec:r Disks}

The disk signals in Horizon-AGN are further explored in this section to shed light on the origin of the negative alignment signals that have been measured in previous works.

Figure \ref{fig:AB DxE HorizonAGN} depicts the $r^2\tilde{\xi}_{g+,2}$ measurements of disks around ellipticals (continuous lines, as in Section \ref{sec:r DxE}) and disks around all galaxies (dashed lines, as in Section \ref{sec:r AB}) in Horizon-AGN.
These measurements are shown for simple (top row, defined by Eq. \ref{eq: inertia tensor}) and reduced shapes (bottom row, defined by Eq. \ref{eq: reduced inertia tensor}) at $z=0$ (left column) and $z=1$ (right column); for the morphological samples as described in previous sections.
As noted previously, the alignment signal amplitude in Horizon-AGN is not very strong, particularly for disks.
Therefore in Figure \ref{fig:AB DxE HorizonAGN}, only the measurement for simple shapes at $z=0$ (top left panel) gives a clear non-zero signal for both types of measurements.
The amplitude diminishes for reduced shapes compared to simple shapes and for $z=1$ when compared to $z=0$.

\begin{figure*}
    \begin{center}
        \centering
    	\includegraphics[width=1.0\textwidth]{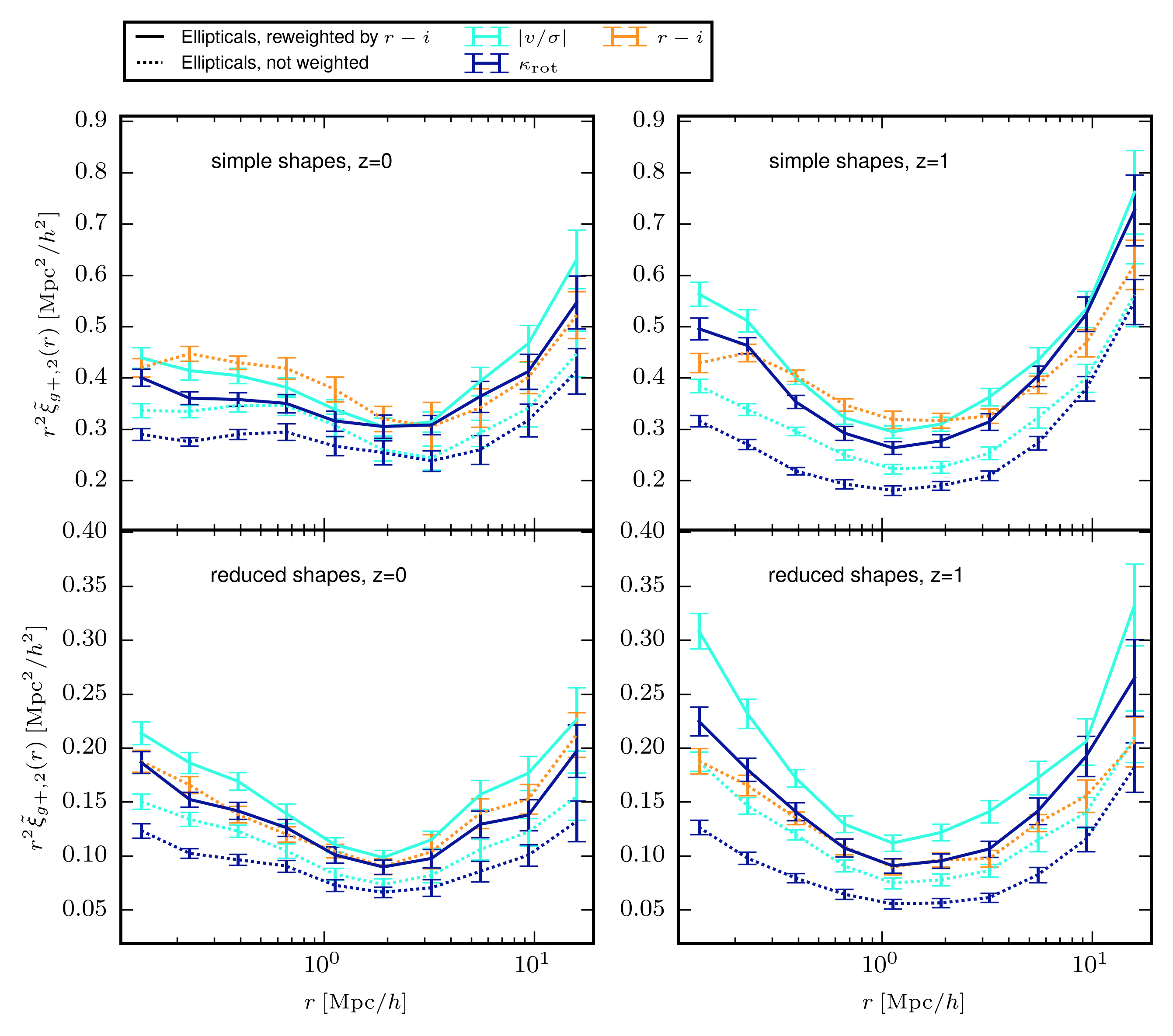}
    	\caption{
        The quadrupoles of ellipticals around all resolved galaxies in TNG300 at $z=0$ (left) and $z=1$ (right) for shapes calculated using the simple (top) or reduced (bottom) inertia tensor.
        The shape samples are defined by the values of the $|v/\sigma|$ (light blue), $\kappa_{\mathrm{rot}}$ (dark blue) or $r-i$ (orange) of the galaxies.
        The dashed lines are the original correlations as in Fig. \ref{fig:AB TNG300}, the continuous lines are the same samples as the dashed lines, but the shape sample is re-weighted to fit the $r-i$ stellar mass distribution (see Figure \ref{fig:Mass distr}).
        The re-weighted correlations show that the stellar mass distributions of the elliptical samples are the driving factor of the IA correlation amplitudes of $|v/\sigma|$ and  $\kappa_{\mathrm{rot}}$.
        }
        \label{fig:AB TNG300 reweighted}
    \end{center}
\end{figure*}

Comparing the IA correlations where the position and shape selection are determined by $|v/\sigma|$ across panels in Figure \ref{fig:AB DxE HorizonAGN}, shows that when measured with reduced shapes at $z=1$, the signal is negative at almost all $r$, while it is positive or consistent with zero for simple shapes and at $z=0$ at all scales. 
This measurement also best matches the choices that led to the negative signal obtained by \cite{Chisari_2016} (see Fig. 14), where the galaxies were split by their $|v/\sigma|$ and a correlation of low $|v/\sigma|$ shapes around high $|v/\sigma|$ positions at $z=0.5$ yielded a negative signal, albeit only at small scales.
The appearance of the negative signal for reduced shapes, not present for the simple shapes, indicates that the cause of this tangential alignment between disks and ellipticals is likely found in the inner regions of galaxies, as these are probed more closely by the reduced shapes.

\subsection{Mass relations}
\label{sec:r Mass}

When selecting a (shape) sample based on a certain variable (e.g. $|v/\sigma|$), it is very difficult to distinguish between the effects of the variable itself, and the effects of its underlying stellar mass distribution on IA.
Based on the assumption that stellar mass correlates heavily with IA amplitude, as shown in many other studies \citep[e.g.][]{Chisari_2015,Tenneti_2016,Samuroff_2021}, the previous sections all support the conclusion that the main trends in amplitude are due to the underlying stellar mass distributions of the sample selections, and secondary effects on small scales are due to (sub-grid) physics.
This section attempts to disentangle these effects by re-weighting the elliptical shape samples in TNG300 defined by $|v/\sigma|$ and $\kappa_{\mathrm{rot}}$ so that their re-weighted stellar mass distributions fit the stellar mass distribution of the $r-i$ elliptical sample.
The weights, $w_j$ in Eq. \ref{eq:S+D}, are defined as follows:
\begin{equation}
    w_j = p_{r-i}/p_{e}
    \label{eq: weights}
\end{equation}
where $p_x$ is the normalised number density of the stellar mass distribution of the elliptical samples defined by $r-i$ or $e \in [|v/\sigma|,\kappa_{\mathrm{rot}}]$.
The stellar mass distributions are binned in logarithmic stellar mass, where the highest mass bin is slightly larger to ensure that there are no bins with too few galaxies.

Recall from Figure \ref{fig:Mass distr}, that the stellar mass distributions of the $|v/\sigma|$ and $\kappa_{\mathrm{rot}}$ ellipticals are very similar and include many low-mass galaxies, whereas the stellar mass distribution of the $r-i$ elliptical sample includes mainly high-mass galaxies.
Due to how the samples are defined (see Section \ref{sec:r distributions}), all three elliptical samples include the same number of galaxies.

Figure \ref{fig:AB TNG300 reweighted} shows the quadrupoles of ellipticals around all galaxies in TNG300 for simple (top panels) and reduced (bottom panels) shapes at $z=0$ (left) and $z=1$ (right), as in Figure \ref{fig:AB TNG300}.
The IA correlations of the elliptical shape samples defined by $|v/\sigma|$ (light blue), $\kappa_{\mathrm{rot}}$ (dark blue) or $r-i$ (orange) are shown in dashed lines.
The continuous lines show the IA correlation functions of the $|v/\sigma|$ and $\kappa_{\mathrm{rot}}$ elliptical samples, when re-weighted according to Eq. \ref{eq: weights}.

Each panel in Figure \ref{fig:AB TNG300 reweighted} shows that when re-weighting the IA correlations of the elliptical shape samples defined by $|v/\sigma|$ (light blue) and $\kappa_{\mathrm{rot}}$ (dark blue), the amplitude of the correlations increases when compared to the non-weighted measurements (dashed).
This supports the hypothesis that stellar mass is the main driver of the intrinsic alignment amplitude.
The effect is robust under shape definition and redshift.
Only for the reduced shapes at $z=1$ (bottom right panel), we see that the ellipticals around galaxy correlation defined by $|v/\sigma|$ hints at possible extra effects of physics, as the correlations differ by 2.66 $\sigma$.
Note that making a similar figure for Horizon-AGN for the ellipticals around galaxies and disks around ellipticals correlations yields a different result.
In that case, the re-weighted $|v/\sigma|$ correlations preserve their sign and amplitude, indicating that stellar mass is not driving the amplitude for this simulation.



\section{Discussion \& Outlook}
\label{sec:discussion and outlook}

This work aims to shed light on the discrepancies between previously measured IA signals of disks in different cosmological hydrodynamical simulations, which vary from positive to negative, depending on the specific simulation, sample selection and measurement methodology.
We explore the alignments of ellipticals (and disks) when measured in TNG300, EAGLE and Horizon-AGN at $z=0$ and $z=1$, for reduced and simple shapes, for various morphological definitions and in multiple sample cross correlations.

Our work goes beyond the existing literature by remeasuring the shapes and variables used in the morphology definitions in a consistent way between simulations.
Although we aim to make a fair comparison between the different simulations, due to the variations in the distributions of the morphologically defining variables, as described in Section \ref{sec:r distributions}, finding a consistent way to define and compare the samples across simulations remains a challenge.

The abundance matching method is, in our opinion, the best way to make the definitions the most consistent across simulations, but does not come without its drawbacks.
Choosing a $90:10$ split does not split the bi-modal colour distribution of TNG300 in its minimum or $\kappa_{\mathrm{rot}}$ and $|v/\sigma|$ according to their theoretically motivated values of $0.5$, $1$, respectively.
Furthermore, the samples chosen at $z=1$ are not the progenitors of the samples at $z=0$.
We do not account for the redshift evolution of the galaxy masses in our mass cut, nor for the evolution of the fraction of disks in our abundance matching fraction.
The current choices are made to reflect what a consistent set of criteria for the selection of a sample yields at different redshifts for different simulations, not to give definitive answers on the redshift evolution of the alignments themselves, although they do provide limited insights in that area.

While this work focusses on the direct comparison of the intrinsic alignment correlations between and within simulations, the origin of the differences between the distributions of the variables and their redshift evolution and between the correlations themselves remains unclear, as underlying stellar mass distributions of the samples, boxsize of the simulation, resolution, sub-grid physics models and galaxy and halo finder algorithms could all be playing a role.
Currently, we are unable to distinguish between physical phenomena and simulation artifacts. 
Without a clear preference from observations or a better understanding of the effects of modelling choices in simulations on these statistics, the interpretation of the results remains limited.
From theory, we do have some expectation on the amplitude of the intrinsic alignment correlation functions, which are given in the works of \citet{Camelio_2015} and \citet{Ghosh_2024} who both estimate the quadrupolar distortion of the phase-space distribution of the stars in a galaxy, created by an external tidal field.
\citet{Ghosh_2024} distinguish between ellipticals and disks, and also estimate how the alignment amplitude of these galaxies depend on properties such as mass, size and S\'ersic-index.
While both studies find lower linear alignment amplitudes than measured in observations, they provide a theoretical, physical motivation for the alignment correlations found in simulations and observations.
At higher order, \citet{Codis15b} attempts to determine the influence of tidal torques on the angular momentum alignment of disk-like galaxies from theory. 
Their findings suggest that below a certain mass threshold, angular momenta align parallel to filaments, and perpendicular above it. 
This could play a role in the interpretation of the sign of the reduced shape alignment signal, although it does not justify the differences found between hydrodynamical simulations.

As mentioned in Section \ref{sec:intro}, making a fair comparison between the IA measurements of disks and ellipticals in hydrodynamical simulations found in literature is challenging due to the vast range of varying measurement choices.
The measured statistics are projected or 3D correlations, or both, or only expressed in terms of the fitted amplitude of the NLA model, $A_{\mathrm{IA}}$, or power spectra, or in the form of a misalignment angle.
For correlations, the alignment between galaxy spins or (semi) major-axis orientation and galaxy positions, filaments or dark matter tracers can be quantified.
The galaxies in the shape and position samples are defined by differing stellar mass cuts and galaxy properties at different redshifts and shape definitions vary.

The IA correlation measurements in TNG300 that \citet{Samuroff_2021} makes are very comparable to this work, as their mass cut is similar to the full shape sample in Section \ref{sec:results}, the shapes are measured using the simple inertia tensor and $r_\mathrm{p}w_{g+}$ is shown at $z=0$ (Figure 5, \citet{Samuroff_2021}).
Comparing this to the disk signal (dashed lines) in Figure \ref{fig:AB TNG300 wg+}, which is the measurement that is most similar, we see that both the shape of the signal and the amplitude ($r_\mathrm{p}w_{g+}\approx0.5$ is very similar between the two works.
The red-blue split, not sub-divided into centrals and satellites, in \citet{Samuroff_2021} is only quantified in the fitted amplitude of IA models, but we can conclude that the amplitude of the red sample is higher than that of the blue, and both samples have a positive amplitude in both works.

The $\mathrm{BTR}$ (Appendix \ref{app: BTR}) and $|v/\sigma|$ splits in this work are similar to the work of \citet{Tenneti_2016}, where both variables are used as a morphological split to measure the ellipticity-direction correlations of disks around ellipticals in MassiveBlack-II \citep{Khandai_2015} and Illustris \citep{Genel2014,Vogelsberger2014}.
These measurements are compared to those in \citet{Chisari_2015}, where $|v/\sigma|$ is used as a morphological proxy to measure the IA correlations of disks around ellipticals in Horizon-AGN.
\citet{Tenneti_2016} measure a positive (radial) disk around elliptical correlation in MassiveBlack-II and Illustris for selections based on both $\mathrm{BTR}$ and $|v/\sigma|$ and show that while the amplitude changes with disk definition, the sign is always positive.

\citet{Chisari_2015} find a persisting tangential projected correlation of disks around ellipticals in Horizon-AGN when using a the reduced inertia tensor (Fig. 14), while it is radial for the simple shapes.
Both \citet{Chisari_2015} and \citet{Tenneti_2016} use similar stellar mass cuts and shape definitions, but vary in redshift, simulation, measurement method and threshold for morphological definition.
Each work differs from this one on multiple points, making a comparison difficult, which is why this work attempts to work with all simulations in a consistent way.
However, in this work we also find persisting radial disk signals for most cases that are studied, with the exception of the reduced disks around ellipticals correlation at $z=1$, when defined by $|v/\sigma|$, which best coincides with the tangential disk alignment measurement in \citet{Chisari_2015}.
Therefore the works are in agreement.

\citet{Delgado_2023} measure $w_{g+}$ in MilleniumTNG, the newer, larger version of TNG300, which is not yet publicly available.
Their selection of disks and ellipticals is based on $\kappa_{\mathrm{rot}}$ and $\mathrm{sSFR}$, following \citet{zhang2023}, and their stellar mass cut is similar to this work.
The scale dependence in $r_{\mathrm{p}}w_{g+}$ and the amplitudes that \citet{Delgado_2023} find in MilleniumTNG (Fig. 3) are similar to those shown in Figure \ref{fig:AB TNG300 wg+} in this work.

There are multiple works that measure the alignments of the spins of disks to filaments or other matter tracers \citep[e.g.][]{welker2017caughtrhythmiicompetitive,Kraljic_2020}.
The relation between galaxy spin and shape is thought to arise due to tidal torquing, as described by tidal torque theory \citep[see][for a review]{schafer_2009}.
The correlation between galaxy spins and shapes in simulations is found to be dependent on morphology, mass, redshift, merger history, and shape type \citep{Chisari_2015,Lee_2022,Moon_2024}, making a one-to-one comparison of spin alignment measurements to this work challenging, although they can offer complementary insights.
In Horizon-AGN, \citet{Chisari_2015} found galaxy spin to be better correlated with the shapes measured by the reduced inertia tensor for intermediate and high $|v/\sigma|$ galaxies, which may offer insight into the origin of the negative IA correlation found for disks when defined by $|v/\sigma|$ for reduced shapes in this work.

\citet{Velliscig_2015} measure the alignments of galaxies in EAGLE via the mean misalignment angle between the major axis of a galaxy and the separation vector to neighbouring subhalos, as a function of separation.
Using the sphericity of the galaxy as a morphological proxy, they find that more spherical galaxies are less aligned, and the signals are only distinguishable at small scales.
This is in line with our measurements in EAGLE, where the disk and elliptical correlations are also only distinct at small scales, and ellipticals are more aligned than disks.

Although this work gives insight into under what circumstances a negative disk signal is measured, further sub-selections of the galaxy samples and larger statistics could help unveil this.  
Measuring the same signal in larger boxes, such as the Horizon Run 5 \citep{Lee_2021}, COLIBRE \citep{schaye2025colibreprojectcosmologicalhydrodynamical,chaikin2025colibrecalibratingsubgridfeedback} and MilleniumTNG \citep{Delgado_2023}, would also be very useful to continue this research.
The effects of boxsize and resolution could also be explored further within these projects, as there will be multiple boxsizes and resolutions available that meet the minimum requirements to measure statistically significant correlations.
To further this research area, future hydrodynamical simulations could focus on obtaining colours that are directly comparable to observations, making it easier to mimic the selections of observational samples.
Furthermore, it would be beneficial if galaxy kinematics are more realistic and understood.

Another area to expand this research to would be the observational realm.
Currently, little is known about how alignments respond to the kinematic properties of galaxies, as the measurement of $|v/\sigma|$ are very difficult to obtain observationally and only available for small samples.
Expanding works like \citet{Pulsoni_2020} that compare kinematic properties of galaxies between hydrodynamical simulations and observations for small samples, will provide valuable insight on the accuracy of the models we currently employ.

This work highlights the importance of consistency when comparing between different disk and elliptical alignments under varying circumstances.
Although hydrodynamical simulations agree on many features of the alignment signal, details like sample selection choices or measurement methodologies can make a substantive difference in outcome. 
Furthermore, this works calls for caution in weak lensing studies that restrict to blue galaxies to mitigate IA effects \citep[e.g.][]{Li21,mccullough2024darkenergysurveyyear}, based on the expectation that blue galaxies do not show significant IA.
This work shows a persisting alignment of blue galaxies in most measurement cases.

\section{Conclusions}
\label{sec:conclusions}
In this work, we have studied the projected correlation functions of the IA of galaxies though measurements of the quadrupole ($\tilde{\xi}_{g+,2}$) in TNG300, EAGLE and Horizon-AGN, at $z=0$ and $z=1$, for simple and reduced shapes, for galaxies split into disks and ellipticals according to their $r-i$, $|v/\sigma|$, $\kappa_{\mathrm{rot}}$ and $\mathrm{BTR}$ (and $M_\star$).
We measured the IA correlations of disks and ellipticals around all resolved galaxies and of disks around ellipticals to gain understanding of the origin of the previously measured disk alignments, which vary from radial to tangential.
Using a consistent comparison across simulations, we are able to shed light on the nature of the disk signal and in which cases it is positive, negative or null.
The main conclusions of this work can be summarised as follows.

\begin{enumerate}
\itemsep0em
    \item The distributions of the $r-i$, $|v/\sigma|$ and $\kappa_{\mathrm{rot}}$ properties of galaxies show variation in shape and redshift evolution across TNG300, EAGLE and Horizon-AGN.
    For $r-i$, the bimodality of the distribution in TNG300, or unimodality in EAGLE and Horizon-AGN is the largest discrepancy. 
    For $|v/\sigma|$ and $\kappa_{\mathrm{rot}}$ it is the opposing redshift evolution between EAGLE and the other two simulations (Fig. \ref{fig:comp sim distr}-\ref{fig:Horizon-AGN distr z}).
    \item The stellar mass distributions of the elliptical samples are fairly consistent across simulations, but vary between morphological selections: where selecting by $|v/\sigma|$ or $\kappa_{\mathrm{rot}}$ include many low-mass galaxies ($M_\star\lesssim10^{10}\mathrm{M}_\odot/h$) and selecting by $r-i$ or $\mathrm{BTR}$ includes mainly higher mass galaxies. (Fig. \ref{fig:Mass distr})
    \item There is consistency between the simulations in the main features of the disks and ellipticals around all resolved galaxies correlations at $z=0$ for simple shapes (Fig. \ref{fig:AB s0}).
        \begin{enumerate}
            \item All elliptical signals are positive; disks are positive or consistent with null.
            \item All ellipticals around galaxies signals have a higher amplitude than the disks around galaxies signals.
            \item The range of amplitudes of the correlations is similar between simulations and morphological definitions within one simulation.
        \end{enumerate}
    \item Differing morphological definitions creates distinct ellipticals around galaxies correlations in TNG300 and Horizon-AGN, where the relative amplitudes are not in the same order between the simulations. (Fig. \ref{fig:AB s0})
    \item Within TNG300, the disks and ellipticals around galaxies correlations are robust across redshift and shape definition, where the amplitude of the reduced shapes is lower than those of simple shapes (Fig. \ref{fig:AB TNG300}).
    \item The IA correlation signal scale dependence for disks around ellipticals diverges for different simulations, although for simple shapes, all signals are positive (TNG300 and EAGLE) or consistent with zero (Horizon-AGN). (See Figs. \ref{fig:DxE s0} and \ref{fig:DxE s1}.)
    \item In Horizon-AGN, both types of disk signals are positive for simple shapes at $z=0$ and null in all other cases, except for the signal of disks around ellipticals when defined by $|v/\sigma|$ for reduced shapes at $z=1$, which is negative. (See Fig. \ref{fig:AB DxE HorizonAGN}.)
    \item When re-weighting the ellipticals around galaxies correlation functions defined by $|v/\sigma|$ or $\kappa_{\mathrm{rot}}$ to match the stellar mass distribution of the $r-i$ elliptical sample, the amplitudes of both of these signals increases compared to the non-weighted versions.
    As the re-weighted correlations agree with the correlation defined by $r-i$, we can conclude that the stellar mass distribution of the shape sample is the main driver of the amplitude of the intrinsic alignment correlation functions. (See Fig. \ref{fig:AB TNG300 reweighted})
\end{enumerate}

\section*{Data availability}
The catalogues with the variables described in Section \ref{sec:md selections} and the galaxy shapes used in this work are publicly available for TNG300\footnote{https://doi.org/10.5281/zenodo.18740718}, EAGLE\footnote{https://doi.org/10.5281/zenodo.18740922} and Horizon-AGN\footnote{https://doi.org/10.5281/zenodo.18740433}.
The correlations can be measured using these catalogues and the public code {\sc{MeasureIA}}\xspace\footnote{https://github.com/MarloesvL/measure\_IA}\footnote{doi:10.5281/zenodo.17252215}.

\section*{Acknowledgements}
    The authors want to thank the research groups of Volker Springel, Joop Schaye and Yohan Dubois for creating the IllustrisTNG, EAGLE and Horizon-AGN simulation data and allowing us to use it in this project.
    This publication is part of the project ``A rising tide: Galaxy intrinsic alignments as a new probe of cosmology and galaxy evolution'' (with project number VI.Vidi.203.011) of the Talent programme Vidi which is (partly) financed by the Dutch Research Council (NWO).

\bibliographystyle{mnras}
\typeout{}
\bibliography{Bibliography}

\begin{appendix}
\input{resolution}
\input{wg+}
\input{ABs0}
\input{BTR}
\end{appendix}

\end{document}

%% file: resolution.tex
\section{Appendix A: Resolution and boxsize}
\label{app:resolution}

\begin{figure*}
    \begin{center}
        \centering
    	\includegraphics[width=1.0\textwidth]{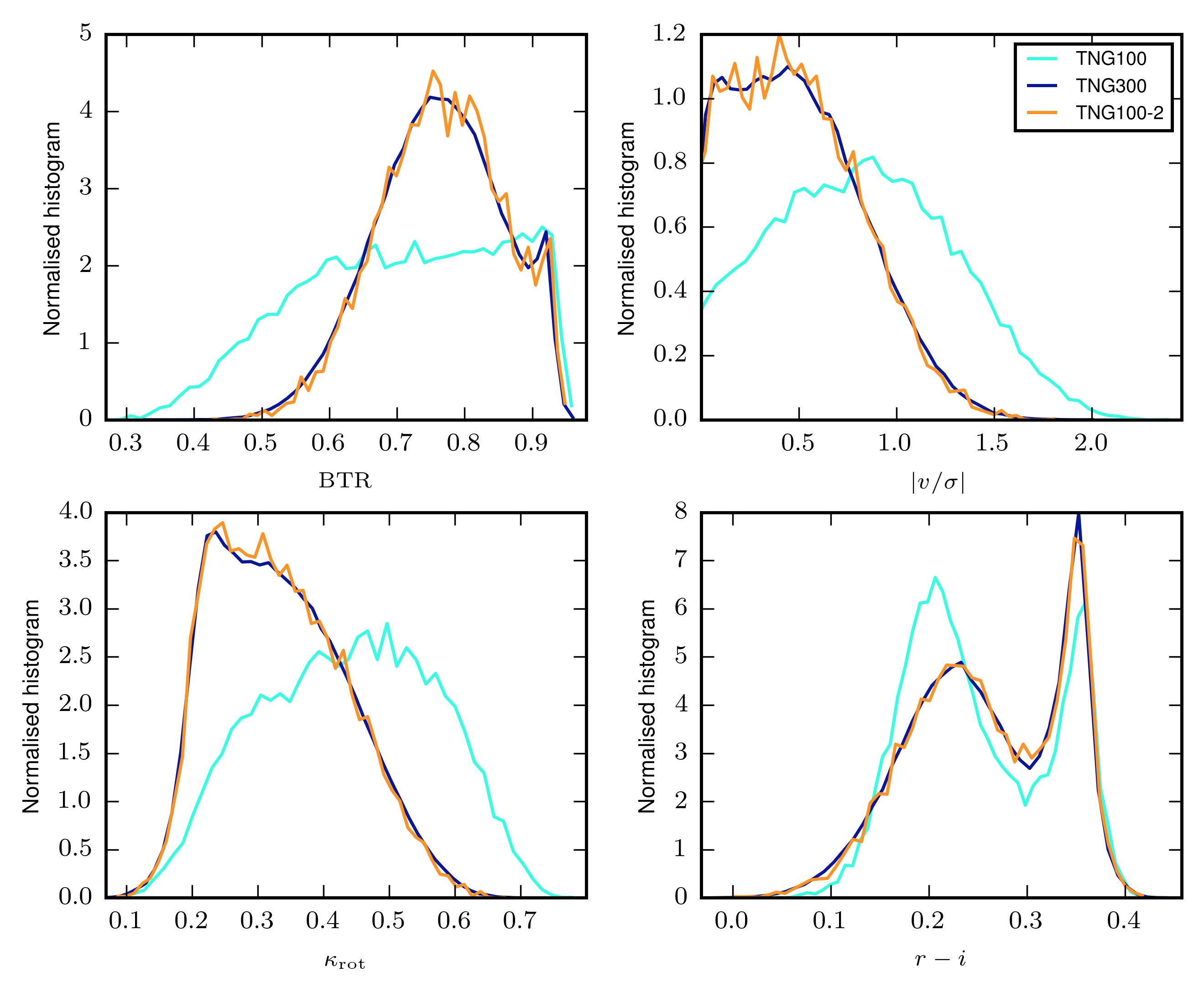}
    	\caption{
        Normalised distributions of BTR (top left panel), $|v/\sigma|$ (top right), $\kappa_{rot}$ (bottom left) and $r-i$ colour (bottom right) in TNG300 (dark blue), TNG100 (light blue) and TNG100-2 (orange).
        Distribution shape comes from the resolution of the simulation and its noisiness from the boxsize.
        }
        \label{fig:Distributions resolution}
    \end{center}
\end{figure*}

When comparing different simulation projects, not only the different models, algorithms and sub-grid physics models can influence the results, but also differences in boxsize and resolution.
Therefore, it is important to explore these effects on both the distributions of the variables and on the intrinsic alignment correlation functions.
This appendix explores these effects within the TNG project, as there are multiple boxsizes and resolutions publicly available.
While this may also hint at the effects of boxsize and resolution in and between the other simulations, the differences in simulation models make it impossible to draw any definitive conclusions about this.
For reference, the simulations are ordered as follows according to their resolution in dark matter particle mass: TNG100$>$EAGLE$>$TNG300$>$Horizon-AGN.
According to boxsize they are ordered as follows: TNG300$>$Horizon-AGN$>$TNG100$>$EAGLE.


Figure \ref{fig:Distributions resolution} shows the normalised distributions of BTR (top left panel), $|v/\sigma|$ (top right), $\kappa_{rot}$ (bottom left) and $r-i$ colour (bottom right) in TNG300 (dark blue), TNG100 (light blue) and TNG100-2 (orange).
Here, TNG300 has a boxsize of $205\,\mathrm{cMpc}/h$ and a dark matter particle mass of $5.9\times10^{7}{\rm M}_\odot$; TNG100 has a boxsize of $75\,\mathrm{cMpc}/h$ and a dark matter particle mass of $7.5\times10^{6}{\rm M}_\odot$ and TNG100-2 has a boxsize of $75\,\mathrm{cMpc}/h$ and a dark matter particle mass of $6.0\times10^{7}{\rm M}_\odot$.
The boxsizes of TNG100 and TNG100-2 are equal, while that of TNG300 is larger, and the resolutions of TNG300 and TNG100-2 are (roughly) equal, while that of TNG100 is higher.

\begin{figure*}
    \begin{center}
        \centering
    	\includegraphics[width=1.0\textwidth]{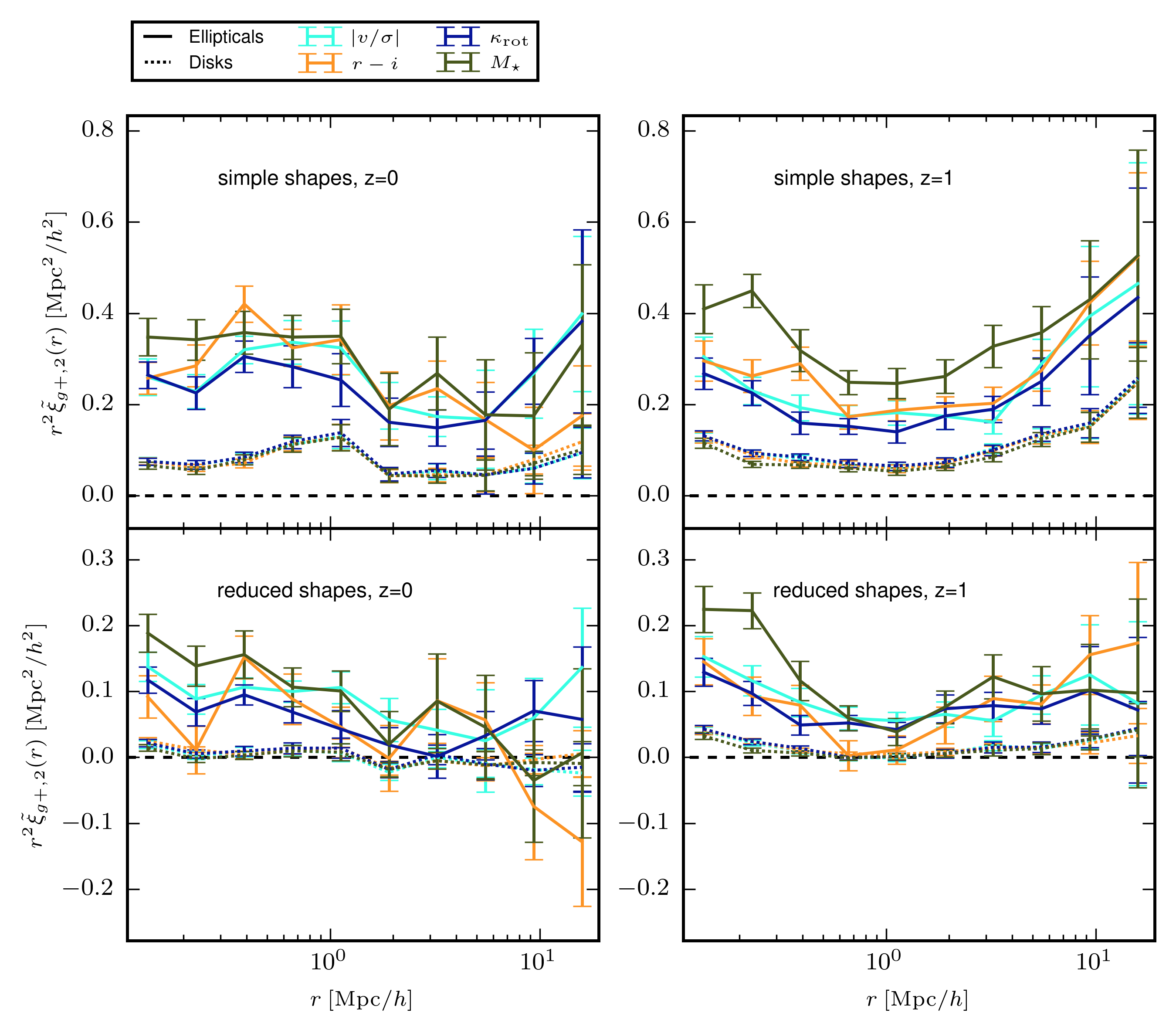}
    	\caption{
        The quadrupoles of ellipticals (continuous lines) and disks (dashed) around all resolved galaxies in TNG100 at $z=0$ (left) and $z=1$ (right) for shapes calculated using the simple (top) or reduced (bottom) inertia tensor.
        The shape samples are defined by the values of the $|v/\sigma|$ (light blue), $\kappa_{rot}$ (dark blue), $r-i$ (orange) or $M_\star$ (green) of the galaxies.
        The results are similar to those of TNG300, given in Fig. \ref{fig:AB TNG300}, albeit noisier.
        }
        \label{fig:AB TNG100}
    \end{center}
\end{figure*}

Figure \ref{fig:Distributions resolution} shows that the TNG100-2 distributions overlap with those of TNG300, indicating a strong impact of resolution on the shape of the distributions and range of values.
The (sub-grid) models determining the distribution shape are often resolution-dependent for simulations.
The TNG100 and TNG100-2 distributions both show the same noisiness, where the TNG300 distributions are smooth, due to the larger sample size in TNG300 when compared to the other two simulations, which is determined by the boxsize.

Figure \ref{fig:AB TNG100} is the same as Figure \ref{fig:AB TNG300}, except for TNG100 instead of TNG300.
It shows the measurements of the quadrupoles of disks (dashed) and ellipticals (continuous) around all galaxies for different morphological definitions.
We created the same figure for TNG100-2 (not shown), which shows very similar results to Figure \ref{fig:AB TNG100}.
Comparing the quadrupole measurements between TNG300, TNG100 and TNG100-2 we can say that the main impact on the results comes from boxsize, which makes the measurements in TNG100(-2) very noisy and does not allow for the same depth of analysis as TNG300.
The overall shapes and amplitudes of the quadrupoles in TNG100(-2) are similar enough to TNG300 and between TNG100 and TNG100-2, that it seems that resolution does not play a significant role here. 

Based on these figures, resolution will likely not influence the results in a significant way.
The caveat here is that simulations from different projects may depend on resolution in a different way, so this appendix can give a definitive answer for TNG300 and only an indication on what is likely for Horizon-AGN and EAGLE.
As for boxsize, the smoothness and the sizes of the error bars in TNG300 indicate that this box is large enough for these measurements.
EAGLE struggles to distinguish between morphological definitions and Horizon-AGN seems to be marginally affected.

%% file: wg+.tex
\section{Appendix B: $w_{g+}$ results}
\label{app:wg+}

\begin{figure*}
    \begin{center}
        \centering
    	\includegraphics[width=1.0\textwidth]{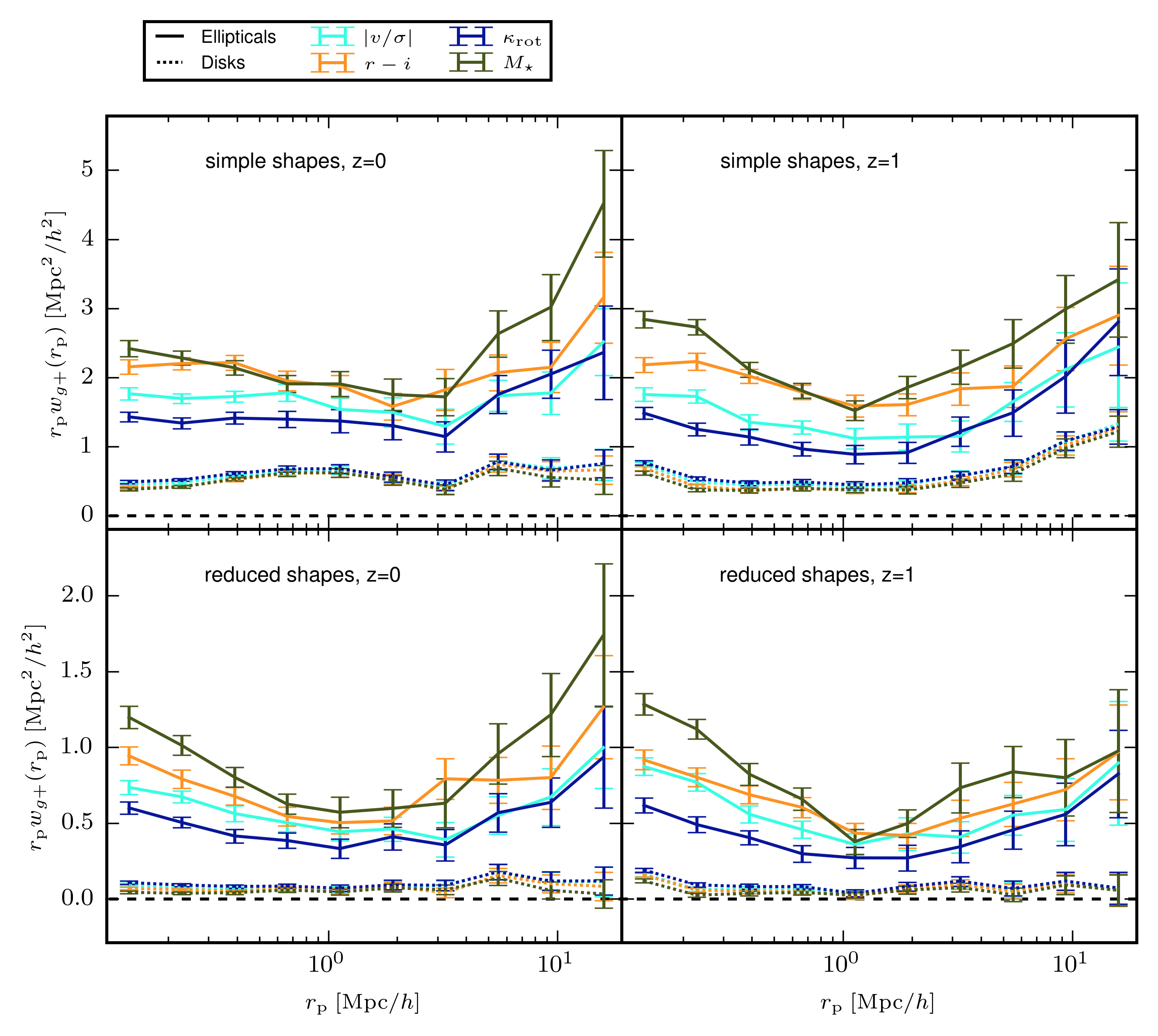}
    	\caption{
        The $r_pw_{g+}$ of ellipticals (continuous lines) and disks (dashed) around all resolved galaxies in TNG300 at $z=0$ (left) and $z=1$ (right) for shapes calculated using the simple (top) or reduced (bottom) inertia tensor.
        The shape samples are defined by the values of the $|v/\sigma|$ (light blue),  $\kappa_{rot}$ (dark blue), $r-i$ (orange) or $M_\star$ (green) of the galaxies.
        The main trends are the same as for the quadrupoles, but the $w_{g+}$ measurements are more noisy.
        }
        \label{fig:AB TNG300 wg+}
    \end{center}
\end{figure*}

This Appendix shows the measurements for $r_p w_{g+}$ in TNG300, to allow for easier comparisons with earlier works.
Any quadrupole measurements shown in this (entire) work are also available for $w_{g+}$ upon request.
The samples in Figure \ref{fig:AB TNG300 wg+} are the same as those in Fig. \ref{fig:AB TNG300}, showing disks (dashed) and ellipticals (continuous) around all galaxies for various morphological definitions, indicated by colour.
As before, the top row shows simple shapes, whereas the bottom row shows reduced shapes, and the results at $z=0$ and $z=1$ are shown in the left and right columns, respectively.

When comparing Figure \ref{fig:AB TNG300 wg+} to Figure \ref{fig:AB TNG300}, we see that the $r_pw_{g+}$ and $r^2\tilde{\xi}_{g+,2}$ produce the same qualitative results.
Due to their definitions, we need to correct for their varying $r_{(p)}$-dependences, to compare 'fairly'.
The relatively scale-independent correlation functions that remain are not identical, but display similar features and amplitudes.
As for the quadrupoles, $w_{g+}$ has a higher amplitude for ellipticals than disks and the amplitudes of the simple shape correlations are higher than for the reduced shapes.
The most notable change is the increase of the error bar sizes and the decreased smoothness of the correlations, resulting in overlapping correlations for the various morphological definitions.
This is also the reason we have chosen to show only the quadrupoles in the main text.

%% file: ABs0.tex
\section{Appendix C: Disks and ellipticals around galaxies: EAGLE and Horizon-AGN}
\label{app:AB EAGLE HorizonAGN}

\begin{figure*}
    \begin{center}
        \centering
    	\includegraphics[angle=90, height=0.95\textheight]{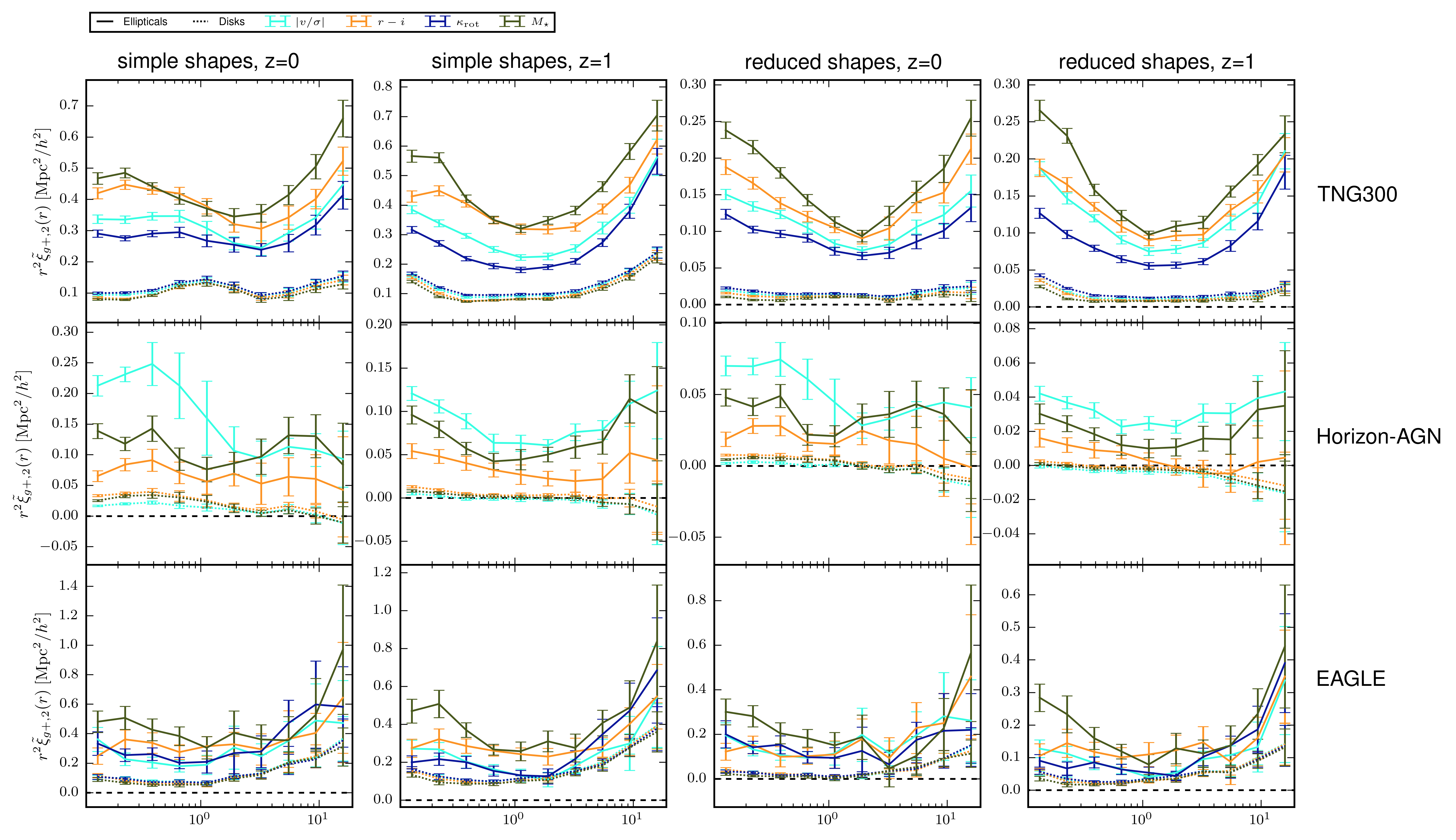}
    	\caption{
        The quadrupoles of disks (dashed) and ellipticals (continuous) around galaxies for various morphological definitions (colour), for simple shapes at $z=0$ (first column), simple shapes at $z=1$ (second column), reduced shapes at $z=0$ (third column), reduced shapes at $z=1$ (fourth column), in EAGLE (bottom row), Horizon-AGN (middle row), and TNG300 (top row, same as Fig. \ref{fig:AB TNG300}).
        The shape samples are defined by the values of the $|v/\sigma|$ (light blue), $\kappa_{rot}$ (dark blue), $r-i$ (orange) or $M_\star$ (green) of the galaxies.
        The scale dependence and ordering of the amplitudes of the different samples persist across redshift and shape definition for each simulation.
        }
        \label{fig:AB all}
    \end{center}
\end{figure*}

This Appendix presents the quadrupoles ($\times r^2$) of disks (dashed) and ellipticals (continuous) around galaxies in Figure \ref{fig:AB all} for various morphological definitions (colour), for simple shapes at $z=0$ (first column), simple shapes at $z=1$ (second column), reduced shapes at $z=0$ (third column), reduced shapes at $z=1$ (fourth column), in EAGLE (bottom row), Horizon-AGN (middle row), and TNG300 (top row), which is the equivalent to Figure \ref{fig:AB TNG300}.

Similar to TNG300, both EAGLE and Horizon-AGN display robustness in their signals across redshift and shape type.
For both EAGLE and Horizon-AGN there is no residual scale dependence in the $r^2\tilde{\xi}_{g+}$ measurements for both shape types at $z=0$ and $z=1$.
Furthermore, elliptical correlation amplitudes are higher than those for the disks.
As for TNG300, in EAGLE and Horizon-AGN the amplitude for the reduced shapes is lower than for the simple shapes, and the amplitudes at $z=0$ and $z=1$ for a single shape type are comparable.

In EAGLE, the sample selections of disks and ellipticals overlap within error bars due to insufficient sample sizes and therefore, not much can be said about the influence of morphological definition on the quadrupoles measured.
In Horizon-AGN, the $|v/\sigma|$ elliptical signal at small scales, $r\lesssim2\mathrm{Mpc}/h$, has a larger amplitude than the $r-i$ elliptical signal, where the offset is larger for simple shapes.
The difference between the simple and reduced shape amplitudes indicates a higher contribution of the outer regions of $|v/\sigma|$ ellipticals to the alignment correlations than the outer regions of $r-i$ ellipticals at 1-halo scales.

%% file: BTR.tex
\section{Appendix D: Bulge-to-total ratio}
\label{app: BTR}
\begin{figure*}
    \begin{center}
        \centering
    	\includegraphics[width=1.0\textwidth]{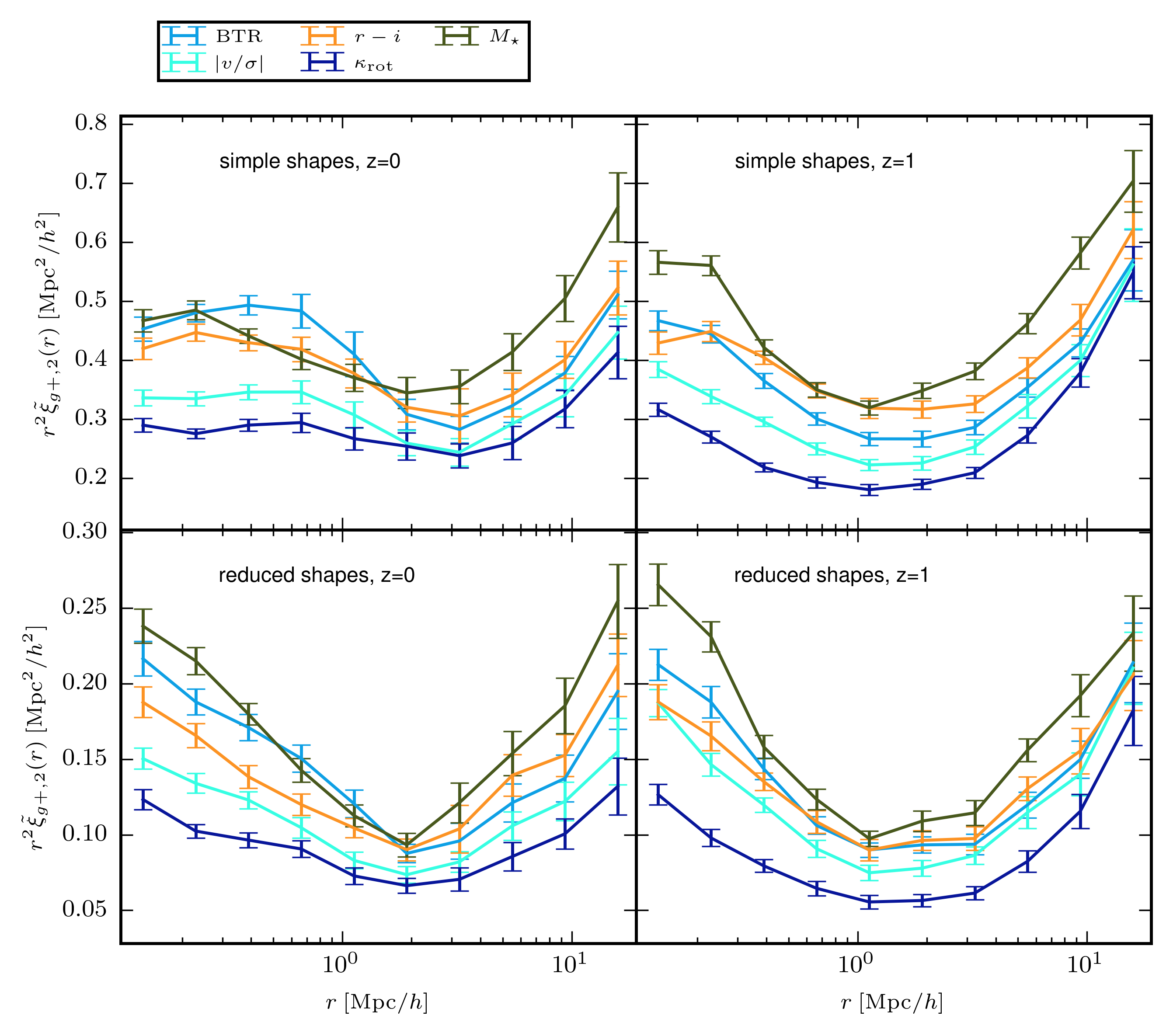}
    	\caption{
        The quadrupoles of ellipticals around all resolved galaxies in TNG300 at $z=0$ (left) and $z=1$ (right) for shapes calculated using the simple (top) or reduced (bottom) inertia tensor.
        The shape samples are defined by the values of the $|v/\sigma|$ (light blue), BTR (medium blue), $\kappa_{rot}$ (dark blue), $r-i$ (orange) or $M_\star$ (green) of the galaxies.
        The BTR elliptical correlations are most comparable to those defined by $r-i$.
        }
        \label{fig:AB TNG300 BTR}
    \end{center}
\end{figure*}

As mentioned in Section \ref{sec:md selections}, the TNG300 simulation provides a catalogue that includes extra kinematic information from which the bulge-to-total ratio (BTR) can be measured.
As we are using a consistent stellar mass cut for our shape sample, and the BTR measurements are only available for galaxies with $M_\star>10^{9.8}\mathrm{M_\odot}/h$, these measurements are discussed separately in this Appendix.
To make the comparison with the other morphological definitions, the elliptical sample has been chosen to be the same size as the other samples, consisting of the galaxies with the highest BTR.
As there are a substantially smaller number of disks left for the BTR sample due to the higher mass cut, the disks around galaxies correlations have been omitted from Figure \ref{fig:AB TNG300 BTR}.
The ellipticals around galaxies correlations shown in Figure \ref{fig:AB TNG300 BTR} for the other morphological variables are the same as in Figure \ref{fig:AB TNG300 BTR} and have been placed for reference.

Figure \ref{fig:AB TNG300 BTR} shows that the IA correlations of BTR ellipticals around galaxies (medium blue) are comparable to the $r-i$ ellipticals around galaxies correlations for both shape types and redshifts.
The two signals overlap within error bars at large scales, $r\gtrsim2\mathrm{Mpc}/h$, which is likely due to their similar stellar mass distributions (Fig. \ref{fig:Mass distr}).
At non-linear, 1-halo scales ($r\lesssim2\mathrm{Mpc}/h$), there are slight differences between the BTR and $r-i$ correlations, indicating that dependence of IA on these properties at small scales are not identical, but no noteworthy deviations are found.